\documentclass{aa}  

\usepackage{graphicx}
\usepackage{txfonts}
\usepackage{natbib}
\usepackage{xcolor}
\bibpunct{(}{)}{;}{a}{}{,} 

\usepackage{placeins}

\usepackage{xpatch}
\usepackage{hyperref}
\hypersetup{
	colorlinks=true,
	breaklinks=true,
	citecolor=blue,
	allcolors=blue,
	frenchlinks=true
}

\makeatletter
\xpatchcmd\NAT@citex
{%
	\@citea\NAT@hyper@{%
		\NAT@nmfmt{\NAT@nm}%
		\hyper@natlinkbreak{\NAT@aysep\NAT@spacechar}{\@citeb\@extra@b@citeb}%
		\NAT@date
	}%
}
{%
	\@citea
	\NAT@nmfmt{\NAT@nm}%
	\NAT@aysep\NAT@spacechar
	\NAT@hyper@{\NAT@date}%
}
{}{}
\xpatchcmd\NAT@citex
{%
	\@citea\NAT@hyper@{%
		\NAT@nmfmt{\NAT@nm}%
		\hyper@natlinkbreak{\NAT@spacechar\NAT@@open\if*#1*\else#1\NAT@spacechar\fi}%
		{\@citeb\@extra@b@citeb}%
		\NAT@date
	}%
}
{
	\@citea
	\NAT@nmfmt{\NAT@nm}%
	\NAT@spacechar\NAT@@open\if*#1*\else#1\NAT@spacechar\fi
	\NAT@hyper@{\NAT@date}%
}
{}{}
\makeatother

\makeatletter
\renewcommand*\aa@pageof{, page \thepage{} of \pageref*{LastPage}}
\makeatother

\begin{document}

   \title{Active galactic nucleus driven jet feedback in cosmologically forming cool-core galaxy clusters I}
   \subtitle{The effect of hierarchical assembly on intra-cluster medium properties}
   \titlerunning{Jets in galaxy clusters I}

   \author{R.~Weinberger
          \inst{1}\thanks{rweinberger@aip.de},
          L. M. Perrone
          \inst{1},
          C. Pfrommer
          \inst{1},
          T. Berlok
          \inst{2},
          E. Puchwein
          \inst{1},
          L. Jlassi
          \inst{1,3},
          R. Talbot
          \inst{4}, \\
          J. Whittingham
          \inst{1},
          R. Pakmor
          \inst{4},
          V. Springel
          \inst{4}
          }
  \authorrunning{R. Weinberger et al.}

   \institute{Leibniz Institute for Astrophysics Potsdam (AIP),
              An der Sternwarte 16, 14482 Potsdam, Germany
              \and
              Niels Bohr Institute, Blegdamsvej 17, DK-2100 Copenhagen, Denmark
              \and
              Institut f\"ur Physik und Astronomie, Universit\"at Potsdam, Karl-Liebknecht-Str. 24/25, 14476 Potsdam, Germany
              \and 
              Max-Planck-Institut f\"ur Astrophysik , Karl-Schwarzschild-Str. 1, 85748 Garching, Germany
             }

   \date{Received September 15, 1996; accepted March 16, 1997}

  \abstract{
  The hot, dilute gaseous atmospheres of cool-core galaxy clusters are excellent probes of astrophysical plasmas at low densities and high temperatures. However, how the interplay between assembly, active galactic nucleus (AGN) feedback and effects of weakly collisional plasmas leads to the observed gas profiles remains uncertain. We study the impact of hierarchical assembly on the intra-cluster medium in cool-core galaxy clusters using hydrodynamic simulations as part of the PICO-Cluster project. We compare cosmological zoom simulations employing an explicit AGN jet model against PICO-Cluster simulations with IllustrisTNG kinetic AGN feedback, as well as against isolated galaxy cluster simulations using the same jet feedback prescription. The stellar and gas fractions of our cosmological galaxy cluster simulations with AGN jet feedback are in excellent agreement with observed galaxy clusters, and the intra-cluster medium (ICM) thermodynamic profiles resemble those of local cool-core galaxy clusters while those run with IllustrisTNG kinetic AGN feedback do not match these observations. In all simulations, cosmological and isolated, the AGN jet heating roughly balances the cooling losses, with star formation being significantly suppressed. The most notable differences between the cosmological and isolated simulations are the resulting velocity and multi-phase structure: gas at radii $> 50\,\mathrm{kpc}$ is shaped by satellite galaxies rather than jet feedback originating form the central galaxy. This leads to significant differences in non-thermal pressure support, with only the cosmological simulations being consistent with recent observations. A second notable difference is the abundance of warm ($<10^5\,\mathrm{K}$) gas beyond the core region, which is absent in our isolated simulation. Overall, our results highlight the need for taking cosmological assembly into account in comparisons of the intra-cluster gas dynamics and its multi-phase nature, while self-regulation is only indirectly altered by hierarchical assembly via merger-driven growth of the central supermassive black hole (SMBH). 
   }


   \maketitle

\section{Introduction}

Galaxy clusters are the largest virialised structures in the Universe. They consist of baryonic and dark matter roughly in the same proportion as the average cosmic fraction \citep{Popesso2024}.
The dominant baryonic mass component is the hot intra-cluster medium (ICM) which fills the volume of the halo and is roughly in hydrostatic equilibrium in the halo's gravitational potential at temperatures of $1\,\mathrm{keV}<kT<10\,\mathrm{keV}$.

Observationally, galaxy clusters are frequently classified into two populations: non-cool-core clusters, which have a centrally less dense, hotter intra-cluster medium with correspondingly larger central cooling time, as opposed to cool-core clusters which have denser, cooler centres, and shorter central cooling times of a few hundred Myr \citep{Cavagnolo2009, Hudson2010, Andrade-Santos2017}. For the latter population of galaxy clusters, the observed amount of cold gas, let alone star formation in the central galaxies is two orders of magnitude lower than cooling-flow estimates from the observed X-ray emission would indicate \citep{Fabian1994, Peterson2003, Peterson2006}. Whether the cold gas is absent or observationally inaccessible is a topic under active investigation \citep{Fabian2022, Fabian2023, Hopkins2025}. 

This lack of star formation can be understood through the presence of X-ray cavities in most cool-core galaxy clusters \citep{Birzan2004, Dunn2006, Hlavacek-Larrondo2015, Olivares2023}.
These cavities coincide with radio emission and are believed to be the remnants of past AGN driven jet activity \citep{Weinberger2017b, Ehlert2018, Jlassi2026, Jlassi2026b}. 
Estimates of the total energy of these jet episodes and the timescales involved indicate that the rate of energy injection via AGN driven jets is comparable to the overall cooling losses in the ICM \citep{Rafferty2006}. This may explain why the gas is not undergoing a cooling flow of several thousands of M$_\odot$~yr$^{-1}$ due to X-ray cooling, at least from a global energetics perspective \citep{McNamara2007, McNamara2012, Fabian2012}.
 
Thermal instability and subsequent star formation, however, are local phenomena, and it is not obvious that a global energy balance leads to a local suppression of star formation. Thus, how the jet energy is distributed throughout the ICM to prevent star formation locally is an open question that is tightly related to the energy dissipation process itself. 
Possible mechanisms include turbulent dissipation \citep{Zhuravleva2014}, turbulent mixing \citep{Kim2003,Sternberg2008}, weak shocks \citep{Yang2016, Li2017}, dissipation of sound waves \citep{Sanders2008}, gravity waves \citep{Reynolds2015, Lehle2026} or cosmic ray heating \citep{Guo2008, Pfrommer2013,Jacob2017, Jacob2017b}. While progress has been made, e.g.~by the detection of shocks in the ICM \citep[][and references therein]{Ubertosi2023, Prunier2025}, direct measurements of ICM velocity dispersion \citep{Russell2026, McNamara2026} as well as in the analysis of hydrodynamical simulations \citep{Martizzi2019, Meenakshi2026}, a definite answer has not yet been reached.

Theoretical studies of galaxy clusters have approached this problem from different ends. From the large-scale, cosmological structure formation perspective, simulating the relatively rare galaxy cluster mass halos requires the modelling of a considerable volume (simulation boxes of ${\sim}300$~Mpc side length or larger) to capture significant enough initial density peaks that produce halos of $10^{15}\,{\rm M}_\odot$. This implies that, even when using zoom-in techniques, the achievable numerical resolution is limited; usually not below kpc scales \citep{Dubois2010, Bahe2017, Barnes2017, Tremmel2019, Bassini2020, Pellissier2023, Nelson2024, Steinwandel2024}. This, in turn, implies that models for AGN feedback applied in this kind of simulation \citep{Booth2009, Ragone-Figueroa2013, Tremmel2017, Weinberger2017, Sullivan2026}
are relatively simple, and generally not able to capture the gradual inflation of hot cavities in the ICM by low-density AGN driven jets \citep[see][for a recent review on numerical models of AGN jets]{Bourne2023}.

In order to study AGN jet feedback in more detail, many works have resorted to setups of isolated, static cluster potentials with an idealized ICM. These are computationally less demanding and consequently allow for higher resolution in the galaxy cluster centre and more accurate AGN jet modelling \citep[e.g.][among others]{Gaspari2011, Li2015, Meece2017, Ehlert2018, Ehlert2023, Beckmann2022, Grete2025, Cammelli2026}.
It should be noted, however, that even among these simpler setups, striking a compromise between runtime and jet modelling fidelity is required. Naturally, the nature of this compromise depends heavily on the addressed scientific question in a particular study. Thus, while of higher physical fidelity than cosmological simulations, the modelling of simulated jets can vary from simulation to simulation, and the precise impact of these different model choices on the feedback distribution and the ICM is not always obvious.

Efforts to combine these two types of simulations have been made \citep{Heinz2006, Mendygral2012, Bourne2021, Yates-Jones2023, Su2026}, 
however, so far these efforts are limited either by the time the jet can be modelled in full detail or by its fidelity. While the resulting simulations capture the effect of cosmological infall on the particular jet outburst, definitively answering questions about self-regulated jet feedback requires Gyr integration times, beyond what is currently possible at the required resolutions.

In this work, we use the AGN-driven jet model established in \citet{Weinberger2017b, Weinberger2023b} and apply it for the first time in a full cosmological context using selected PICO-Cluster cosmological zoom initial conditions \citep{Berlok2026}. We study the effects of cosmological assembly and long-term evolution, and compare to the isolated galaxy cluster setup of \citet{Ehlert2023}, adjusted for resolutions that are within reach of cosmological simulations, and to the reference cosmological zoom simulations run with the IllustrisTNG model \citep{Weinberger2017, Pillepich2018} presented in \citet{Berlok2026} to assess the importance of AGN feedback modelling.

The paper is structured as follows: In Section~\ref{sec:method}, we describe the simulation setup and employed models. We present the results in Section~\ref{sec:result}, discuss them in Section~\ref{sec:discussion}, and conclude in Section~\ref{sec:conclusion}.

\section{Methods}
\label{sec:method}

We perform cosmological zoom simulations of three massive galaxy clusters ($M_{200}\approx 10^{15}\,{\rm M}_\odot$) in a $\Lambda$ cold dark matter universe, assuming cosmological parameters consistent with the \citet{PlanckCollaboration2020}, i.e.\ with $\Omega_{\rm m,0} = 0.32$, $\Omega_\Lambda = 0.68$, $\Omega_{\rm b,0}=0.049$, $h=0.67$, $\sigma_8=0.81$ and $n_\textrm{s}=0.97$. We compare these cosmological simulations to hydrodynamic simulations of an isolated galaxy cluster setup run for $14\,{\rm Gyr}$ \citep[initial conditions  and setup from][]{Ehlert2023}.

The cosmological zoom simulations start with initial conditions at redshift $z=127$ and follow the evolution of dark matter and stars by modelling them as collisionless particles that are subject to gravity. For the gas, we compute ideal hydrodynamics in an expanding spacetime, taking radiative cooling and heating into account and regularizing the collapse of gas using an ISM and star formation model based on an effective equation of state \citep{Springel2003}. This is done with the \textsc{Arepo} code \citep{Springel2010, Pakmor2016, Weinberger2020}, using a tree particle-mesh method for the collisionless $N$-body system and a finite-volume approach on a moving, unstructured Voronoi mesh for the hydrodynamics. Our fiducial hydrodynamics model employs AGN jet feedback \citep{Weinberger2017b,Weinberger2023b} as well as galaxy formation physics adopted from the IllustrisTNG simulations \citep{Weinberger2017, Pillepich2018}, with some modifications in the AGN jet feedback model, which we discuss below. Those cosmological simulations will be compared to the magneto-hydrodynamics PICO-Cluster simulations \citep{Berlok2026} run with IllustrisTNG galaxy formation physics, which employs kinetic AGN feedback at late times \citep{Weinberger2017}.

\subsection{Galaxy formation model}

Radiative cooling is modelled as in IllustrisTNG, taking into account primordial and metal-line cooling down to $10^4\,{\rm K}$ \citep{Katz1996, Wiersma2009}, as well as using a spatially uniform UV background from \citet{Faucher-Giguere2009}, self-shielding corrections from \citet{Rahmati2013} and AGN radiation fields as described in \citet{Vogelsberger2013}.

Gas denser than a threshold value of $\rho=2.33\times10^{-25}$~g~cm$^{-3}$ (a hydrogen number density $n_{\rm H} = 0.1\,{\rm cm}^{-3}$) can either be on an effective equation of state where it forms stars according to a calibrated Schmidt-like star formation law \citep{Springel2003, Vogelsberger2013} or hotter than the respective effective temperature. As in IllustrisTNG, we use a softer equation of state compared to \citet{Springel2003} by using a linear combination between the original model and an isothermal component at $10^4\,{\rm K}$. 
Stars are evolved assuming a Chabrier IMF. Mass return, including chemical enrichment, is modelled assuming single stellar populations per simulation star particle \citep{Pillepich2018}.

On top of the pressurization of the ISM modelled via an effective equation of state, stellar feedback is assumed to launch galactic winds. These are modelled using temporarily hydrodynamically decoupled wind particles that are spawned stochastically from star forming cells \citep{Springel2003, Vogelsberger2013}, with velocity and metal-loading scalings as used in IllustrisTNG \citep{Pillepich2018}. 

Super-massive black holes (SMBHs) are seeded with mass $8\times10^{5}\,h^{-1}{\rm M}_\odot$ in every halo exceeding $5\times10^{10}\,h^{-1}{\rm M}_\odot$ that does not yet contain a SMBH. This is done by frequently running a friend-of-friends halo finding algorithm on the fly and using the resulting group as a measure of halo mass. Black holes are repositioned to the galactic potential minimum, with mergers between SMBHs that come within a distance of one softening length happening instantaneously \citep{Weinberger2017}. 

Accretion onto black holes is modelled using the \citet{Bondi1952} formula, estimating the gas density and sound speed from the surroundings of a black hole in a kernel-weighted fashion using 64 neighbouring cells and limiting the accretion rate to the Eddington accretion rate.
Feedback from AGNs is injected in two mutually exclusive feedback modes. In case of high, sub-Eddington accretion rates\footnote{We do not allow for super-Eddington accretion in these simulations.}, AGN feedback happens via continuous thermal energy deposition in the SMBH surroundings with a rate
\begin{align}
    \dot{E} = \epsilon_\mathrm{f} \epsilon_\mathrm{r} \dot{M} c^2,
\end{align}
where $\epsilon_\mathrm{r}=0.2$ is the radiative efficiency, $\epsilon_\mathrm{f}=0.1$ the feedback efficiency in the thermal mode, $\dot{M}$ the accretion rate, and $c$ the vacuum speed of light. At low accretion rates, AGN feedback is modelled in the form of continuous jets, which will be described in detail in the following. The dividing threshold accretion rate between the two modes, $\dot{M}_\text{thresh}$, follows the IllustrisTNG parameterization \citep{Weinberger2017}, with a minor modification: for SMBHs less massive than $10^7\,{\rm M}_\odot$, we assume thermal feedback independent of accretion rate (IllustrisTNG does not have such a cutoff and low-mass black holes are able to transition to kinetic feedback given sufficiently low accretion rates). The main reason for this are resolution limitations of the employed jet model, which results in an inability to adequately model the comparably low-power jets originating from these low-mass black holes. In practice, we would expect these low-power jets to remain confined or interact substantially within the central kpc \citep[e.g.][]{Mukherjee2016, Borodina2025}. We thus set the dividing line between thermal and jet AGN feedback as
\begin{align}
    \frac{\dot{M}_\text{thresh}}{\dot{M}_\text{Edd}} = 
    \begin{cases}
    \min\left[ 0.002 \left(\frac{M}{10^8\,\mathrm{M}_\odot}\right)^2, 0.1 \right] \, &\text{for} \, M \geq 10^{7}\,\text{M}_\odot \\
    0 &\text{for} \, M<10^7\,\text{M}_\odot.
    \end{cases}
\end{align}
At $z=0$, averaged over the last $8$ snapshots, we find between $70$ and $90$ SMBHs per simulated galaxy cluster,  of which between $4$ and $6$ AGNs are in jet mode.

\subsection{AGN jet feedback model}

The jet injection routine originates from the one presented in \citet{Weinberger2017b},  modified to fit the needs of self-regulated setups, i.e., with an accretion rate estimate that is coupled to the jet power as presented in \citet{Weinberger2023b}. Specifically, the jet power is
\begin{align}
    \dot{E} = \epsilon_\mathrm{\rm f, jet} \dot{M} c^2,
\end{align}
with the jet feedback efficiency, $\epsilon_\mathrm{\rm f, jet} = 0.2$ matching the kinetic feedback efficiency in the IllustrisTNG model \citep{Weinberger2017}. Note that this is significantly larger than the efficiency used in \citet{Ehlert2023}, who use $\epsilon_\mathrm{\rm f, jet} = 0.01$ as a fiducial value.

In the surroundings of each SMBH particle, we define an inner spherical region which we refer to as the jet launching region, as well as an outer spherical shell which we refer to as the surroundings, or the accretion region. The inner jet launching region is split into two half-spheres cut by the $y-z$ coordinate plane of the simulation box at the $x$-position of the black hole particles. Jets are launched in the positive and negative $x$-directions away from the black hole particle. Note that we do not vary the jet direction, and neither employ an opening angle nor jet precession. This is justified by our previous findings that our fiducial jet quickly isotropizes in a galaxy cluster centre upon interacting with cold filaments and clouds \citep{Ehlert2023}.

If there is sufficient available AGN jet energy, we establish the jet’s thermodynamic state by assigning it a (proper) density of $\rho_\mathrm{jet} = 10^{-28}\,{\rm g\,cm}^{-3}$, and set its thermal energy so that its pressure matches that of the surrounding environment. To ensure we conserve momentum exactly while also ensuring both directions get the same amount of kinetic energy, we require that the mass contained within the two half-spheres of the jet is the same. As a result, the actual density used may differ slightly from the target value if the jet launching cells have differing volumes, which can happen with a moving, unstructured mesh. 

Setting up the jet state resembles prescribing boundary conditions, but since it operates on live cells, it removes mass from the simulation. This drained mass is added to the black hole particle as gravitating mass, where it subsequently serves as a reservoir for accretion. To account for the removed mass, we reduce the thermal energy in the jet injection cells by
\begin{align}
    \Delta E_{\rm mass} = \Delta m \left\langle u \right\rangle,
\end{align}
where $\Delta m$ is the drained mass and $\left\langle u \right\rangle$ is the mass-weighted specific internal energy of the surrounding gas. This correction is applied only when mass is removed from the cells, i.e., when $\Delta m > 0$. We then subtract the energy required to bring the jet cells into pressure equilibrium with their surroundings from the total available energy budget. The remaining energy is injected as kinetic kicks into the jet cells in a kernel-weighted fashion.

This scheme produces qualitatively distinct outcomes depending on the jet power. In the low-power jet regime, where little or no kinetic energy remains after establishing pressure equilibrium, the algorithm creates small, underdense cavities in the vicinity of the SMBH. In the high-power jet regime, kinetic energy dominates and the routine drives a supersonic, collimated jet, consistent with resolved jet simulations \citep{Weinberger2023b}.

The specifics of the jet routine ensure an average jet density and precise energy injection, yet other parameters of interest such as its velocity at injection are variable. For the main periods of interest, the central jet has luminosities that are close to the cooling rate of the halo, with around $10^{44}-10^{45}$~erg~s$^{-1}$. Using the cross-section of the injection radius as jet area, the resulting jet velocities for a kinetic jet energy of $10^{44}$~erg~s$^{-1}$ would be around $5\times10^4$~km~s$^{-1}$, i.e., in the mildly relativistic regime. Notably, we do not see cases of sustained relativistic jets in these simulations, justifiying the use of non-relativistic hydrodynamics.

Unlike isolated setups with a single central SMBH, there are many SMBHs present in cosmological simulations, thus jet regions of two black holes can share cells. While this is a very specific, short-lived configuration, it might lead to spurious behaviours if left unchecked. Thus, if a cell has already been used as a jet cell of another jet, the second jet is prevented from being active at the same time, and its available feedback energy is kept for future jet injection events. Which SMBH is the first and which the second is determined by the sorting of particles along a Peano-Hilbert curve intrinsic to the code, and without particular physical meaning.

Another peculiarity of cosmological simulations are extreme resolution and timestep gradients between cells. In order to avoid creating inconsistent hydrodynamical states in the time-integration, we ensure all cells in the surroundings of the black hole are on the same timestep, defined by the minimum timestep criterion of the involved cells.
We also enforce a refinement criterion around the jet region, making sure the jet region is sampled with cells\footnote{I.e., the mesh-generating point of at least one cell is in each jet half-sphere.} at all times. 

To mark where the jet was injected, we use two passive scalars. These scalars follow the flow of the gas and trace the jet material over time. We use the first scalar as an additional refinement criterion, reducing the target mass to
\begin{align}
    m_\mathrm{target} = \min( V_\mathrm{jet, target} \rho X_\mathrm{jet}^{-1}, m_\mathrm{target,0} ),
\end{align}
with a target volume of $V_\textrm{jet, target}^{1/3} = 1.39 \,\mathrm{kpc}$ and a continuous transition to an ICM target mass, $m_\mathrm{target,0}=10^7\,\mathrm{M}_\odot$ as the jet mass fraction $X_\mathrm{jet}$ decreases. Since $X_\mathrm{jet} \leq 1$, this implies that only cells with gas densities $<2.5\times10^{-25}\,\mathrm{g}\,\mathrm{cm}^{-3}$ can be subject to the additional jet refinement. Cells with higher density, most notably star-forming cells, all have approximately the same target mass, leading to  star particles being formed out of these cells also having approximately the same mass. 
\textsc{Arepo} refines (de-refines) cells with masses above twice (below half) this target mass. In order to not get a pileup of highly resolved jet cells from past outbursts, we decay the scalar exponentially with a characteristic lifetime of $1$~Gyr, i.e. long enough to capture the feedback-relevant jet and lobe dynamics, but short enough to reduce the number of high resolution cells. 
The second jet scalar, in contrast, does not decay and is preserved once injected. This second scalar is used when discussing the fate of the jet plasma in a companion paper.

A few further modifications compared to \citet{Weinberger2023b} are included to increase the numerical stability and to handle the extreme resolution gradients inherent in cosmological simulations. First, a maximum (physical) jet injection radius is chosen to be $0.5\,h^{-1}$~kpc in order to avoid unrealistically large jet regions. Second, a specific energy limiter is applied, where the jet is only injected until the point where the specific energy in the injection region, kinetic or thermal, is below the vacuum speed of light squared, $c^2$. Cases of such extreme instantaneous jet power can occur if dense gas clumps enter the accretion region. If such cases occur, we simply inject the energy up to a specific energy of $c^2$ in the jet region and save the rest of the available energy for subsequent timesteps. While this limiter greatly increases numerical robustness, in practice it is rarely triggered, and its practical implications on the state of the cluster are minor, with resulting cluster properties being close to indistinguishable between runs with and without it.

The spatial resolution of the jet in the cosmological simulations is about a factor of two below the fiducial resolution simulations presented in \citet{Ehlert2023}, while the fiducial jet resolution of \citet{Ehlert2023} is used for the isolated simulation. As shown in the resolution study of \citet{Weinberger2023b}, these resolutions lead to far from converged results for the propagation of the jet. We note, however, that the resulting impact on the surrounding ICM and its thermodynamic profiles are far less sensitive to the resolution in the simulation. Appendix~\ref{sec:appendix_res} discusses resolution effects on thermodynamic profiles using an isolated halo (the resolution of the presented cosmological simulation correspond to the low-resolution runs).

\subsection{Initial conditions}

The simulations are part of the PICO-Cluster project (Plasmas In COsmological Clusters, \citealt{Berlok2026}), using halos number $239$, $260$ and $420$ (subsequently referred to as cluster A, cluster B and cluster C, respectively) with cluster masses at redshift $z=0$ of $M_{200}=\{1.071,1.149,0.885\}\times10^{15}\,{\rm M}_\odot$ and $M_{500}=\{0.728,0.681,0.530\}\times10^{15}\,{\rm M}_\odot$ in the original PICO-Cluster simulation with zoom level 12. Note that throughout this paper we measure overdensities in halos relative to the critical density of the Universe.
The halos have been selected from a $1\,h^{-1}{\rm Gpc}$ side length non-radiative magnetohydrodynamical + $N$-body parent simulation with $1024^3$ simulation particles, and zoomed initial conditions were created with the \textsc{CosmoZoomIC} code (Puchwein et al., in prep.). Halos of interest were resimulated at different resolutions; zoom level 12 corresponds to $12^3$ times the parent mass resolution. 

For this study, we selected the three halos of around $10^{15}\,{\rm M}_\odot$ that showed the lowest central entropy in runs performed with the (unmodified) IllustrisTNG model \citep{Weinberger2017, Pillepich2018}. This was done to avoid simulating non-cool-core galaxy clusters with computationally expensive AGN feedback in systems where it plays a minor role at $z=0$. We note however that this choice comes at the cost of not having a clear selection function of the objects presented in this work.
In the (high-resolution) regions where the jet does not dictate the refinement, the target gas resolution is $1.1\times10^7\,{\rm M}_\odot$ and the gravitational softening length is $4.8$ comoving kpc for $z\ge1$ and $2.4$ proper kpc for $z<1$. Gas particles have a comparable softening length, however adjusted to their cell size and at least $0.6$ comoving kpc.

\begin{figure*}[ht]
    \includegraphics{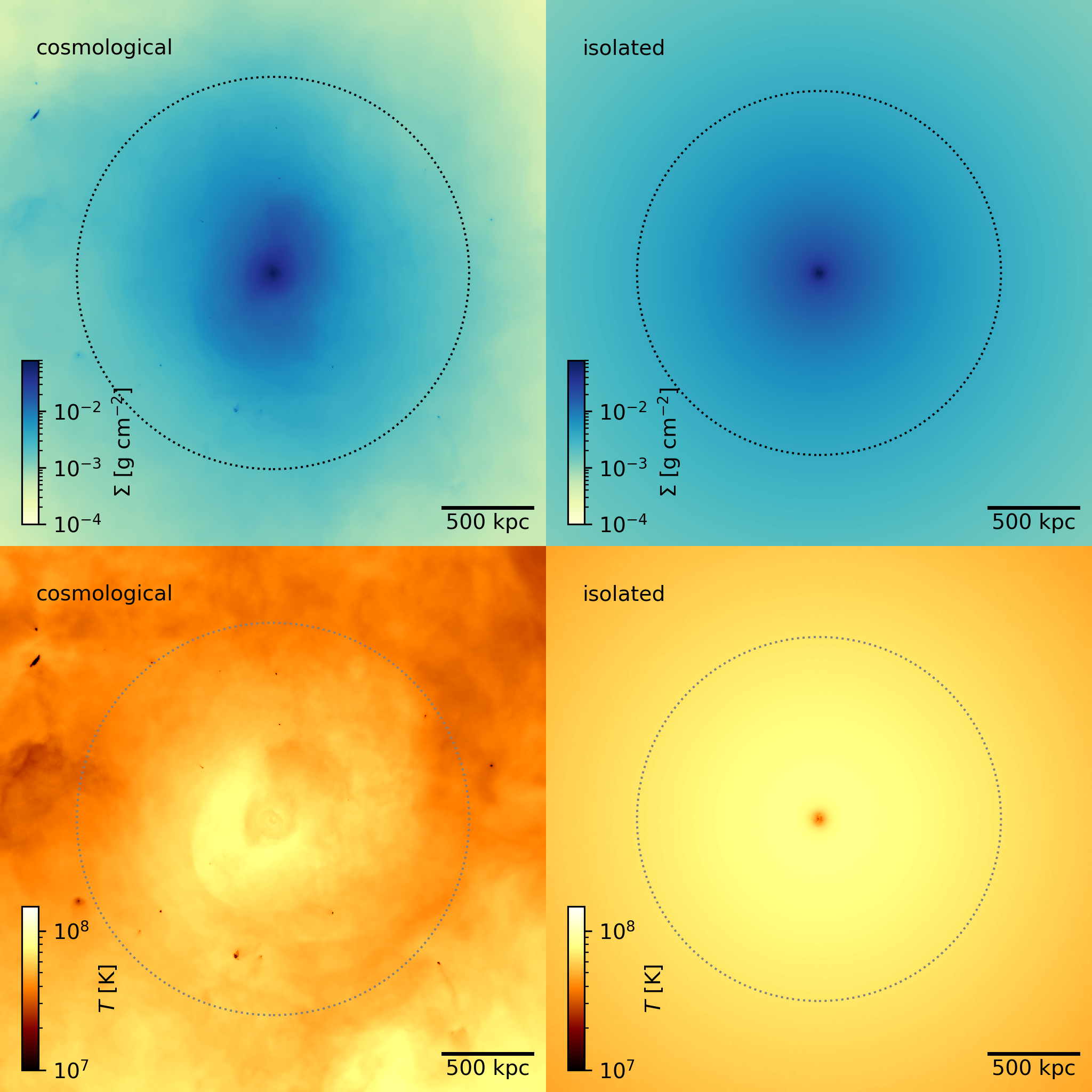}
    \caption{Column density (top) and mass-weighted temperature (bottom) projections of the ICM of a cosmologically forming halo at $z=0$ (left) compared to a corresponding isolated halo after $1\,\mathrm{Gyr}$ (right). The projection depth is $3\,{\rm Mpc}$. The dotted circle indicates $0.5\,R_{200}$. While the overall distribution in the gaseous halo is similar, the isolated setup shows fewer perturbations and a larger degree of spherical symmetry, likely due to the lack of a cosmological formation history through mergers.}
    \label{fig:projections}
\end{figure*}

We compare the cosmological simulations to isolated galaxy cluster simulations that use the initial conditions of \citet{Ehlert2023}. These use an analytic gravitational potential with a NFW cluster potential with mass $8\times 10^{14}\,{\rm M}_\odot$, radius $2\,{\rm Mpc}$, concentration parameter $5$ and a central galaxy potential modelled as an isothermal sphere with mass $2.4\times10^{11}\,{\rm M}_\odot$ and radius $15\,{\rm kpc}$. A SMBH of mass $3\times 10^{9}\,h^{-1}{\rm M}_\odot$ resides in the center of the potential. The ICM is set up with a density profile adopted from a fit to the Perseus galaxy cluster \citep{Churazov2003}, rescaled to $h=0.67$,
\begin{align}
    n_{\rm e} =~~& 46 \times 10^{-3} \left[ 1 + \left(\frac{r}{60\,{\rm kpc}})^2\right)\right]^{-1.8} {\rm cm}^{-3} \\
    &+ 4.7\times 10^{-3} \left[ 1 + \left( \frac{r}{240\,{\rm kpc}} \right)^2 \right]^{-0.87} {\rm cm}^{-3},
\end{align}
with the temperature chosen to bring the ICM in hydrostatic equilibrium with the analytic potential. Self-gravity is neglected. Unlike \citet{Ehlert2023}, we do not include magnetic fields in the simulations, which is achieved by removing it from the original initial conditions, for simplicity. Velocity and isobaric temperature fluctuations are introduced, with fluctuation following power spectrum matching a Kolmogorov slope on scales smaller than $(37.5\,{\rm kpc})^{-1}$ and  white noise on larger scales.  We note that these fluctuations are introduced to ensure the system gradually settles into a steady-state rather than triggering an outburst at a characteristic time due to artificial spherical symmetry of the initial conditions. Throughout this paper, we only analyse the simulation after an initialization period of $1\,{\rm Gyr}$, i.e., once the steady-state has been reached and the initialized turbulence has largely decayed.

\section{Results}
\label{sec:result}

\subsection{Galaxy cluster baryonic content}

\begin{figure*}[ht]
    \includegraphics[]{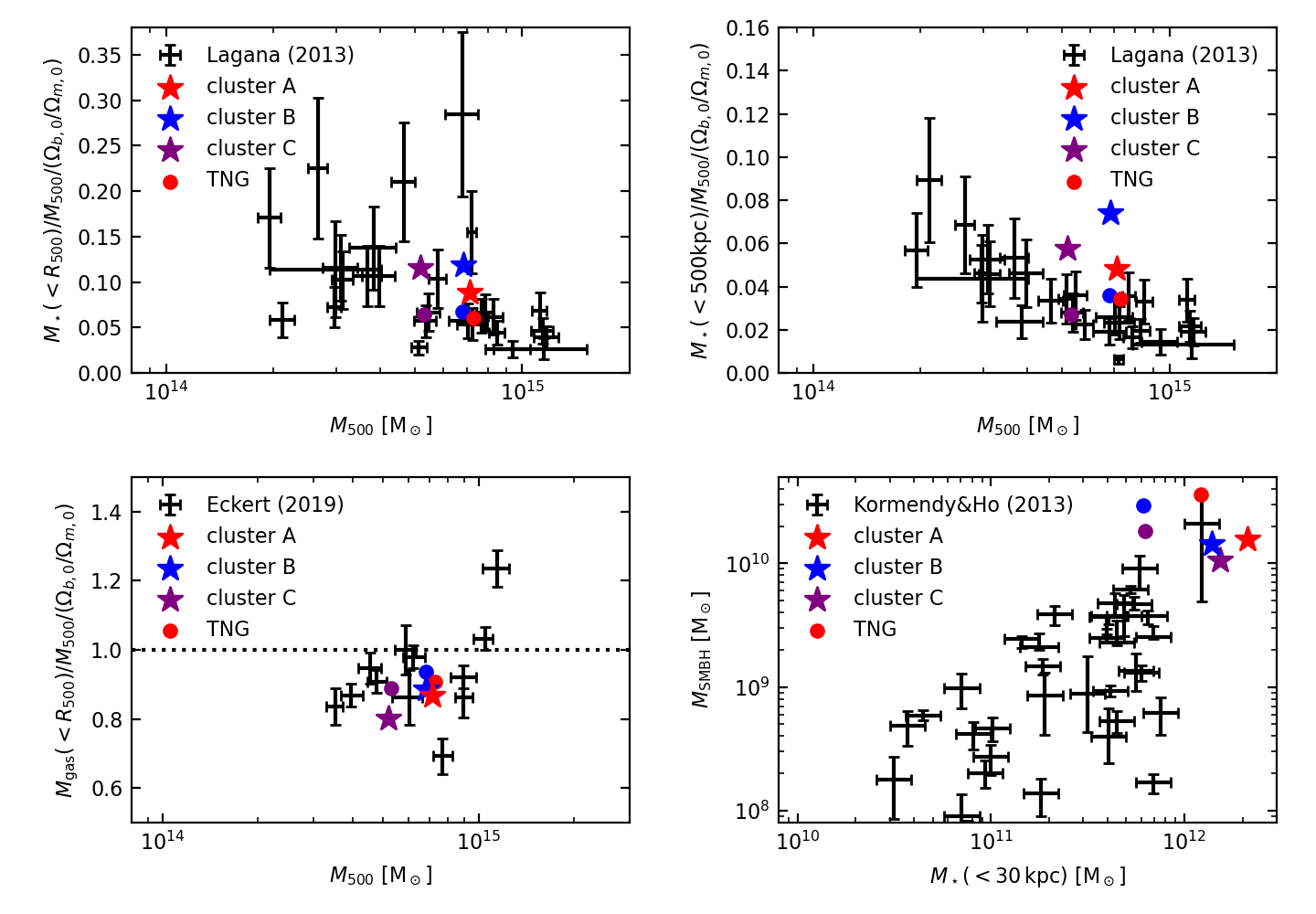}
    \caption{Baryonic content of the simulated cosmological zoom-in halos at $z=0$. Top left: Stellar mass fraction within $R_{500}$ vs. halo mass. The stellar content is in good agreement with data from \citet{Lagana2013}. Top right: Stellar mass fraction within $500\,\mathrm{kpc}$ vs. halo mass. The data from \citet{Lagana2013} represents the stellar mass fractions within $R_{2500}$, i.e., for a similar aperture. Bottom left: Gas mass fraction vs. halo mass within $R_{500}$. The data points are observed galaxy clusters \citep{Eckert2019}. Bottom right: Black hole mass vs. stellar mass within $30\,\mathrm{kpc}$. The data points are measurements from local elliptical galaxies compiled by \citet{Kormendy2013}. The filled circles are the respective values for the PICO-Cluster halos simulated with the IllustrisTNG model.}
    \label{fig:masses_multiplot}
\end{figure*}

Figure~\ref{fig:projections} shows the gas column density (top) and the mass-weighted temperature (bottom) of the ICM in a cosmological halo (cluster~A in left panels; the projections of the other two clusters look comparable) and in the isolated halo (right panels). The cosmologically forming halo is shown at $z=0$, the isolated halo after $1$~Gyr of evolution, i.e., after it has reached a self-regulated equilibrium \citep{Ehlert2023}. The dotted circles indicates $0.5\,R_{200} \approx 1\,\mathrm{Mpc}$. The resulting gaseous halos are relatively similar, with the cosmological halo showing substructure and non spherically symmetric features, while the isolated halo is generally smoother and has a more pronounced drop in central temperature.

Figure~\ref{fig:masses_multiplot} shows the stellar mass fraction within $R_{500}$ vs. $M_{500}$ (top left), the stellar mass fraction within $500$~kpc vs. $M_{500}$ (top right), the gas mass fraction vs. $M_{500}$ (bottom left) and the black hole mass vs.~stellar mass within $30$~kpc (bottom right) relations. The stars indicate the runs with the AGN jet model, while the circles indicate the respective PICO-Cluster simulations run with the fiducial IllustrisTNG model. While comparing three simulated, non-representative systems to observations is clearly insufficient to assess the quality of the model in detail, it is nonetheless important to check that the global gas, stellar and black hole properties are in agreement with low redshift galaxy clusters.
The three cosmologically forming halos have masses of $5\times 10^{14}\,\mathrm{M}_\odot\leq M_{500} \leq 8\times10^{14}\,\mathrm{M}_\odot$ and gas mass fractions of about $80\% - 90\%$ of the cosmic value, in excellent agreement with observationally inferred gas mass fractions of galaxy clusters with comparable mass \citep{Eckert2019}. The stellar mass fraction within $R_{500}$ lies around $10\%$ of the cosmic baryon fraction for all three halos, putting the total baryon fraction at around $90\% - 100\%$. These stellar fractions are in excellent agreement with observed systems \citep{Lagana2013}. Within the smaller aperture of $500\,\mathrm{kpc}$, roughly corresponding to $R_{2500}$, the stellar mass is in slight excess by a factor $1.5 - 2$ to both the observed values and the simulations of the same halos with the IllustrisTNG model. 
The simulated clusters contain black holes with masses in the range $10^{10}\,\mathrm{M}_\odot \leq M_\bullet \leq 1.5\times 10^{10}\,\mathrm{M}_\odot$, putting them in the same range as the most massive black holes with dynamical mass measurements in the local Universe \citep{Kormendy2013}. Notably, the respective SMBH masses of the IllustrisTNG clusters are systematically larger. In summary, the baryonic content of our simulated galaxy clusters is in excellent agreement with observations globally, with a potentially slightly overmassive central galaxy. In the following, we proceed to investigate the detailed structure of the simulated galaxy clusters' cores.

\begin{figure*}[ht]
    \includegraphics[]{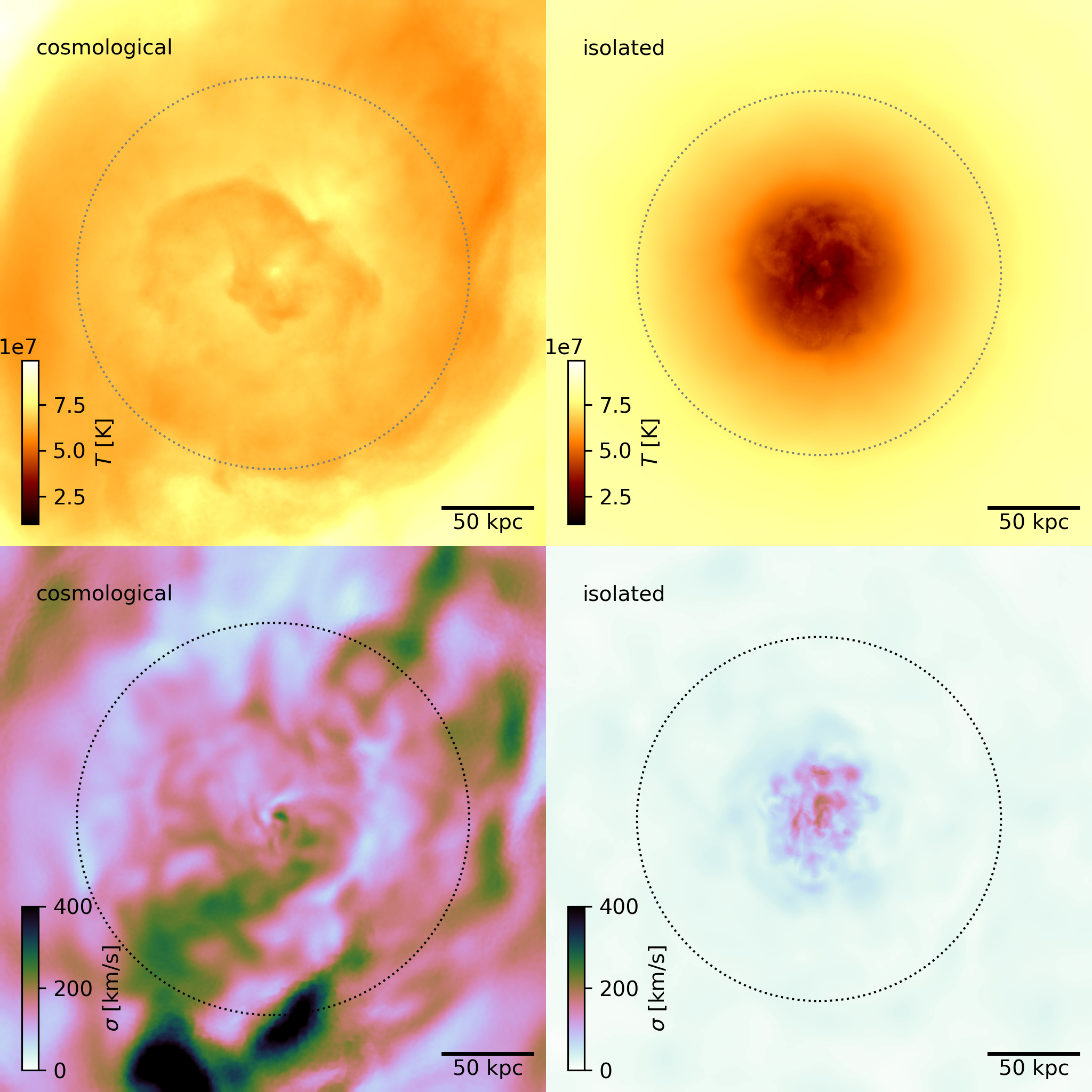}
    \caption{Mass-weighted temperature (top) and line-of-sight velocity dispersion (bottom) projections of the ICM of a cosmologically forming halo at $z=0$ (left) compared to the isolated halo after $1\,\mathrm{Gyr}$ (right). The projection depth is $300\,{\rm kpc}$. The dotted circle indicates $0.05\,R_{200}\approx 100\,\mathrm{kpc}$. While the temperature projections are a zoom-in from Fig.~\ref{fig:projections}, the line of sight velocity dispersion indicates a very different behaviour outside several tens of kpc due to the cosmological formation history.}
    \label{fig:projections_core}
\end{figure*}

Figure~\ref{fig:projections_core} shows mass weighted temperature (top) and line-of-sight velocity dispersion (bottom) projections, this time restricted to the core with a spatial extent and projection depth of $300\,\mathrm{kpc}$, with the dotted circle marking the radius $0.05\,R_{200} \approx 100\,\mathrm{kpc}$. The differences in the temperature, which is in practice a zoom-in on the bottom panel of Fig.~\ref{fig:projections}, highlights the more pronounced cool core in the isolated simulation. The cosmologically forming halo, in contrast, exhibits a more extended, perturbed and less pronounced temperature drop. Most striking, however, are the differences in the line-of-sight velocity dispersion maps. While the cosmological halo exhibits local variations in the range from $50\,\mathrm{km}\,\mathrm{s}^{-1}$ to $400\,\mathrm{km}\,\mathrm{s}^{-1}$ reminiscent of turbulence, no clear radial dependence is discernable. The isolated halo shows similar velocity dispersion levels of $100\,\mathrm{km}\,\mathrm{s}^{-1}$ to $200\,\mathrm{km}\,\mathrm{s}^{-1}$ in the centre, but a rapid drop-off at radii of a few tens of kpc, with practically no velocity dispersion at larger radii. 

\subsection{The driver of intra-cluster medium turbulence}

\begin{figure}[ht]
    \includegraphics[]{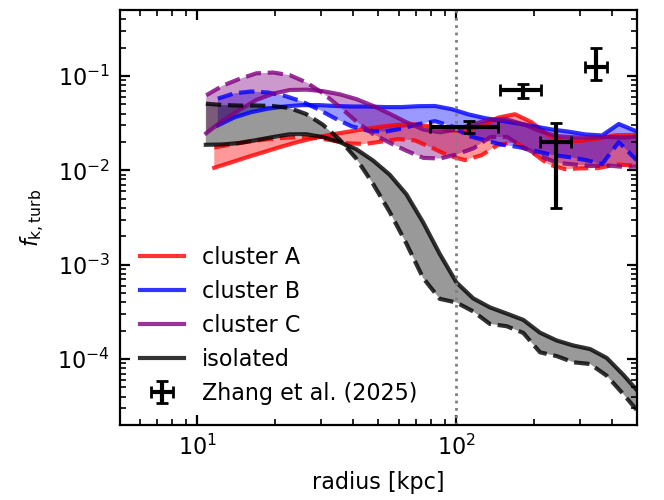}
    \caption{Non-thermal pressure fraction $f_\mathrm{k,turb}$ of the X-ray emitting intra-cluster gas as a function of radius for the four different simulations. Dashed (solid) lines correspond to an effective filter spread of $\sigma_\ell = 20$~kpc ($\sigma_\ell = 40$~kpc), with the shaded region in between. For comparison, we include the recent measurements of the Perseus cluster \citep{Zhang2025}. The turbulence at the levels measured by XRISM cannot be attributed to our central AGN driven jets that are responsible for self-regulation. }
    \label{fig:nt_profiles}
\end{figure}

To quantify the difference in velocities, Fig.~\ref{fig:nt_profiles} shows the profile of the non-thermal pressure fraction, applying the spatial filtering of \citet{Perrone2026}, which makes use of the Paicos analysis package \citep{Berlok2024}, to isolate turbulent from bulk flows:
\begin{align}\label{eq:fkturb}
    f_\mathrm{k,turb} = 
    \frac{\mathcal{E}_\mathrm{k,turb} }{\mathcal{E}_\mathrm{k,turb}+E_\mathrm{th}}=
    \frac{\mathcal{M}^2_\mathrm{1D}}{\mathcal{M}^2_\mathrm{1D} + \gamma^{-1}},
\end{align}
where $\mathcal{E}_\mathrm{k,turb}$ is the volume-integrated turbulent energy below a filtering scale $\ell$, $E_\mathrm{th}$ is the thermal energy, $\gamma = 5/3$ is the adiabatic index and $\mathcal{M}_\mathrm{1D}$ denotes the one-dimensional Mach number defined via the turbulent velocity dispersion obtained through spatial filtering. Note that for the second equality to hold, we assume subsonic, homogenous, isotropic turbulence. 
We use a Gaussian kernel for the spatial filtering and we choose two different filter scales $\ell$ to quantify the scale dependence of the turbulent energy. In terms of their effective spread $\sigma_\ell$ \citep[the variance of the Gaussian kernel, see][]{Perrone2026}, they correspond to 
$\sigma_\ell=20\,\mathrm{kpc}$ (dashed lines in Fig.~\ref{fig:nt_profiles}) and $\sigma_\ell=40\,\mathrm{kpc}$ (solid lines).

We compare the values to recent XRISM observations of the Perseus cluster \citep{Zhang2025}. The cosmologically forming halos show levels of non-thermal pressure similar to the observed values, while the isolated halo shows a steep decline in non-thermal pressure in the outskirts, clearly discrepant from the observed levels of gas motion. We note that the filtering scales chosen here are on the lower end compared to \citet{Perrone2026}. This is necessary in order to properly capture the levels of turbulence in the centre of the isolated halo, which would otherwise appear in the radial profiles of Fig.~\ref{fig:nt_profiles} lower and more extended simply by virtue of spatial filtering. For a detailed mock XRISM observation, we refer the reader to \citet{Bellomi2026}. Independent of measurement details, however, it is evident that the measured levels of gas motion at radii $> 50$~kpc cannot be produced by the central AGN jet, but are a consequence of cosmological infall or interaction processes of infalling galaxies with the ICM ($30\%-60\%$ of the jet power originates from non-central SMBHs at $z=0$). At smaller radii, however, the non-thermal pressure fraction is similar for isolated and cosmologically forming halos. This indicates a common, internal driving process, likely the AGN jet heating-cooling cycle which we will focus on in the following.

\subsection{Self-regulated heating-cooling cycle}

\begin{figure}[ht]
    \includegraphics[]{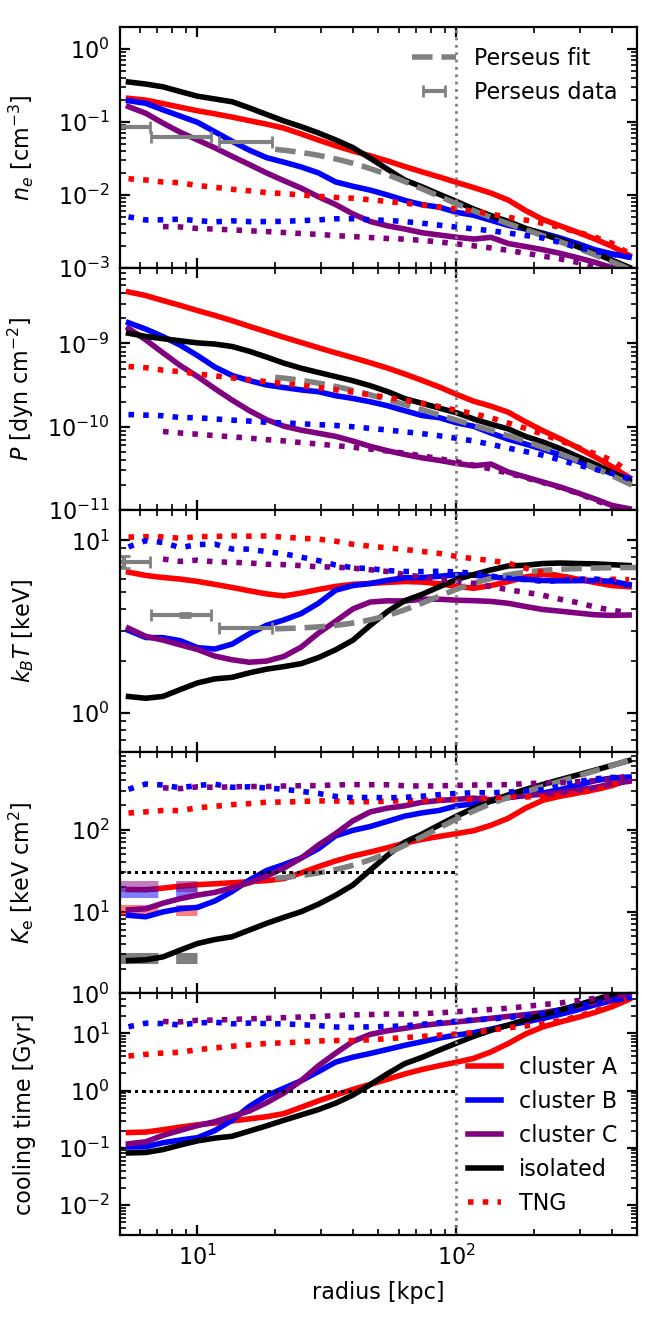}
    \caption{Top to bottom: electron number density, pressure, temperature, entropy and cooling time profiles of halos at $z=0$ (coloured solid lines), for the isolated halo after $1\,\mathrm{Gyr}$ (black line). The coloured dotted lines indicate the profiles of the respective simulations using the IllustrisTNG model. See text for the emission-weighting procedure. The vertical dotted line indicates the radius $0.05\,R_{200} \approx 100\,{\rm kpc}$. The horizontal dotted lines in entropy and cooling time denote commonly used cool-core -- non-cool-core divide. For number density and temperature, we show the Perseus galaxy cluster fit and central data \citep{Churazov2003}. The thick dashed lines in the centre corresponds to the equilibrium entropy of the \citet{Weinberger2026} model of the respective simulation. Both isolated and cosmological AGN jet simulations develop cool-cores. In contrast, the IllustrisTNG model produces non-cool-core states in the same halos.}
    \label{fig:profiles}
\end{figure}

We show the profiles of electron number density, thermal pressure, temperature, cluster entropy and cooling time (top to bottom) in Figure~\ref{fig:profiles}. As explained in \citet{Berlok2026}, we weigh the temperature profile with the bremsstrahlung emission in the X-ray band $(0.5$--$9)\,{\rm keV}$ and reconstruct the density profile from the spherical X-ray emission profile and the emission-weighted temperature profile (assuming that bremsstrahlung dominates the emission for these massive clusters). The pressure, entropy and cooling-time profiles are then obtained from these two thermodynamic profiles. The red, blue and purple lines show the three cosmologically forming halos. Gas in gravitationally bound subhalos was excluded from the profile. The dotted line marks the radius $0.05\,R_{200}$ of the isolated halo (though the other halos have similar core radii). The horizontal lines in the entropy panel at $K_{\rm e}=30$~keV~cm$^{2}$ and in the cooling time panel at $1$~Gyr are empirical divides of cool-core and non-cool-core galaxy clusters \citep{Barnes2018}. We also show the electron density, temperature and cluster entropy fit to observations of the Perseus galaxy cluster \citep[both fit and data from][]{Churazov2003}, scaled to a Hubble constant of $h=0.67$. The fit only matches the observed data at radii $\geq30\,\mathrm{kpc}$. Because of this, we show the data at lower radius directly in the temperature and electron number density plots.

Using the central entropy or central cooling time criterion, all simulated clusters with AGN jet feedback end up in a cool-core state by redshift $z=0$. This is in contrast to the PICO-Cluster simulations performed with the IllustrisTNG model, which produces non-cool-core clusters for these halos. Note that the cosmologically forming halos were selected to have the lowest central entropy in a parent sample, thus hinting at a fundamental shortcoming of the IllustrisTNG model at these mass scales \citep[see also][]{Berlok2026}. For the simulations using jets, variations between the different halos and the isolated setup are of comparable magnitude. Most notably, however, the isolated halo is the densest in the central $20$~kpc, has a substantially more pronounced temperature drop, and consequently features gas with lower entropy and shorter cooling time in the core. While these profiles are taken at a given snapshot, we note that the thermodynamic profiles of the intra-cluster gas in this isolated setup do not vary substantially with time \citep{Ehlert2023}, even after $10\,\mathrm{Gyr}$ simulation time (see Appendix~\ref{sec:appendix_res}). 

\begin{figure}[ht]
    \includegraphics[]{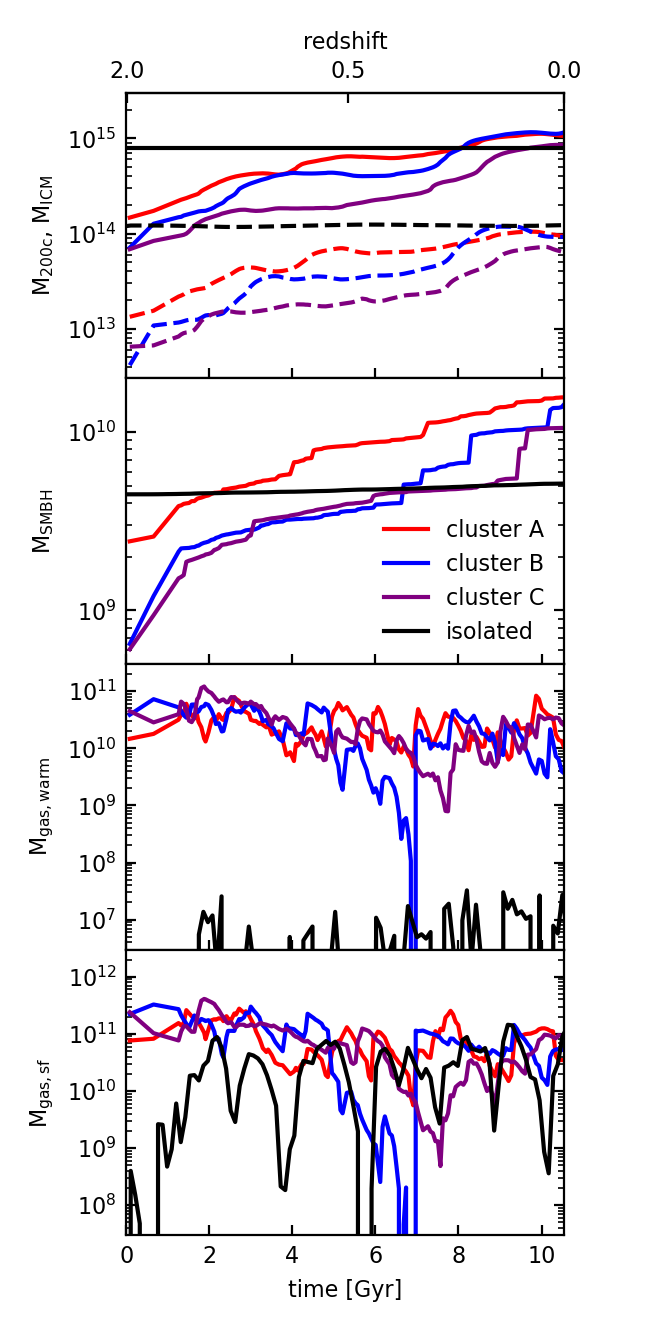}
    \caption{Top to bottom: time evolution of cluster mass (solid) and intra-cluster gas mass within $R_{500}$ (dashed), SMBH mass, warm ($T<10^{5}\,\mathrm{K}$) gas mass and star forming gas mass. The top $x$-axis indicates the corresponding redshift for the cosmological simulations. Note that $t=0$ is arbitrarily identified with $z=2$ for the cosmological simulations, and after a $1\,\mathrm{Gyr}$ initialization period for the isolated simulation. Halo, ICM and black hole mass growth in the cosmological halos is substantial. Warm, non star-forming gas is almost absent in the isolated simulation.}
    \label{fig:mass_time}
\end{figure}

While it is tempting to attribute the steeper profile in the core to the isolated nature of the simulations, and a flattening to a hydrodynamical core to the effect of infalling substructure, there is a more subtle effect at play: Fig.~\ref{fig:mass_time} shows the halo mass (top panel, solid line) and most massive SMBH mass (second from top) as a function of time. While the halo mass in the isolated simulation is kept constant by construction, the cosmological halos only grow to a comparable mass toward $z=0$. The black holes, importantly, outgrow the one in the isolated simulations, reaching masses between $1-1.5\times10^{10}\,\mathrm{M}_\odot$, while the isolated black hole has a final mass of around $5\times 10^{9}\,\mathrm{M}_\odot$. The mass growth of these high mass SMBHs takes place mostly though mergers \citep{Weinberger2018}, manifesting themselves as jumps in the SMBH mass-time diagram that are absent in isolated simulations. Growth through accretion contributes only around $5\times10^8\,\mathrm{M}_\odot$, translating to an average SMBH growth rate of about $4\times 10^{-2}\,\mathrm{M}_\odot\,\mathrm{yr}^{-1}$.

As recently pointed out \citep{Weinberger2026}, and previously discussed in \citet{Cattaneo2006}, the central entropy is tightly linked to the mass of the central SMBH. To test this idea quantitatively, we apply the equilibrium entropy model of \citet{Weinberger2026},
\begin{align}
    K_{\rm e} = 14.5\,{\rm keV}\,{\rm cm}^2 \left( \frac{\epsilon}{0.2} \right)^{2/3} \left( \frac{M_\bullet}{10^{10}\,{\rm M}_\odot} \right)^{4/3} \left( \frac{L_{\rm cool}}{10^{45}\,{\rm erg}\,{s}^{-1}} \right)^{-2/3}, \label{eq:entropy}
\end{align}
where $\epsilon$ is the jet efficiency, $M_\bullet$ the SMBH mass and $L_{\rm cool}$ the ICM cooling luminosity (measured within $R_{200}$).
We show the results as dashed horizontal lines in the centre of the entropy profile in Fig.~\ref{fig:profiles}. The model correctly predicts central entropies of the cosmological simulations between $10$~keV~cm$^{2}$ and $20$~keV~cm$^{2}$. The prediction for the isolated simulation is an order of magnitude reduced. Since the cooling luminosities are comparable, the observed differences in central cluster entropy are consistent with originating from the difference in SMBH mass growth history between cosmological and isolated simulations.

\begin{figure*}[ht]
    \includegraphics[]{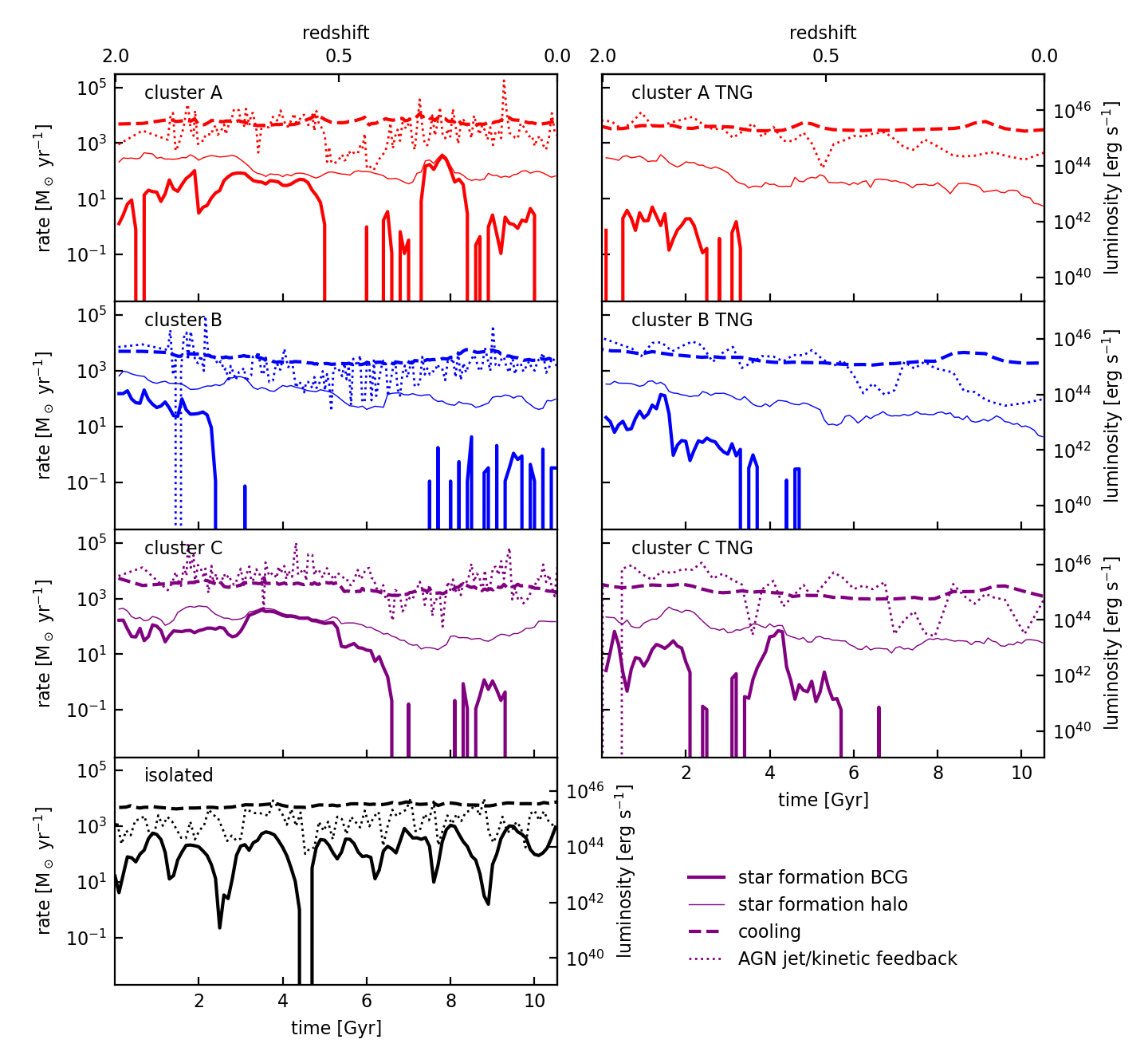}
    \caption{In-situ star formation of the brightest cluster galaxy (BCG, within $30\,{\rm kpc}$; thick solid) and the entire halo (within $R_{200}$; thin solid), cooling rate (dashed) and AGN jet feedback rate (thin dotted) normalized by the virial specific energy of the simulated clusters. The right $y$-axis shows the corresponding luminosity for $z=0$ masses (energy rate per unit time assuming the respective virial specific energy). Note that $t=0$ is arbitrarily identified with $z=2$ for the cosmological simulations, and after a $1\,\mathrm{Gyr}$ initialization period for the isolated simulation. Energetically, star formation in the BCG is only a residual of a roughly balanced heating and cooling rate. }
    \label{fig:sfr_energy_rate_time}
\end{figure*}

We have established that all the clusters simulated with the AGN jet model have cool cores and substantial losses via radiative cooling. We now look at the global energetics of the clusters in Fig.~\ref{fig:sfr_energy_rate_time}. The different rows show the different halos, the solid line shows the in-situ star formation rate, the thick line denotes that of the central galaxy (located within a sphere of $<30\,{\rm kpc}$), the thin line of the entire halo. The left column denotes the simulations using the AGN jet model, the right column the corresponding PICO-Cluster simulations using the fiducial IllustrisTNG model. Cosmological halos start out with relatively high star formation rate of $100\,\mathrm{M}_\odot\,\mathrm{yr}^{-1}$ at time zero (which we set at redshift 2). At some later time, however, the in-situ star formation in the central galaxy drops, in the case of halos~2 and 3 substantially, to a very small amount of residual star formation in the central galaxy. As indicated by the thin line, this does not mean that there is no in-situ star formation in the halo, since some of the cluster galaxies do have ongoing star formation. The isolated halo experiences a periodic increase in star formation followed by a shorter period of low star formation, with remarkable regularity of around $1.5\,\mathrm{Gyr}$ per cycle. The variability timescale indicates that the feature is linked to the global state of the gas (sound crossing time of order $t_\mathrm{s} \approx 1\,\textrm{Mpc}/1000\,\textrm{km}\,\textrm{s}^{-1} \approx 1\,\mathrm{Gyr}$), but is notably absent in the runs that consider cosmological assembly. Interestingly, however, the levels of star formation in the halo of the cosmologically forming halos are comparable to the levels of star formation in the core of the isolated galaxy cluster. The simulations using the IllustrisTNG model show lower overall star formation rates, most visibly in the central galaxy, consistent with their overall lower stellar mass fraction at $z=0$.

The thick dashed and thin dotted lines in Fig.~\ref{fig:sfr_energy_rate_time} represent the cooling rate within $R_{200}$ and AGN jet heating rate, calculated as the respective luminosities, $L = \dot{m}\, u_\mathrm{vir}$, where $\dot{m}$ is the mass (cooling or jet heating) rate and $u_\mathrm{vir} = 0.5\,G\, M_{200}\, R_{200}^{-1}$ denotes the virial specific energy. The resulting cooling flow rate and heating rates are substantially higher than the star formation rate, between $10^3\,\mathrm{M}_\odot\,\mathrm{yr}^{-1}$ and $10^4\,\mathrm{M}_\odot\,\mathrm{yr}^{-1}$. Unlike the cooling rate, which is relatively steady, the feedback rate varies on relatively short timescales (and likely also on timescales shorter than the time between outputs). Importantly, averages of heating and cooling approximately match in all, cosmological and isolated halos. Overall, the comparison of cooling, heating and star formation in Fig.~\ref{fig:sfr_energy_rate_time} emphasises the nature of the ICM after cosmic noon of being in a cooling-heating balance when averaged over long enough timescales with some residual star formation. While this basic picture holds for both isolated and cosmological setups, details like intrinsic (long-duration) duty cycles that establish in isolation seem to be erased by cosmological accretion. Unlike star formation, the cooling and heating rates do not change substantially between jet and IllustrisTNG model. Note that the seemingly lower time variability of the IllustrisTNG simulations are merely a consequence of a lower simulation snapshot frequency and not intrinsic to the model.

\subsection{Phase distribution and the origin of warm gas}
 
\begin{figure*}[ht]
    \includegraphics[]{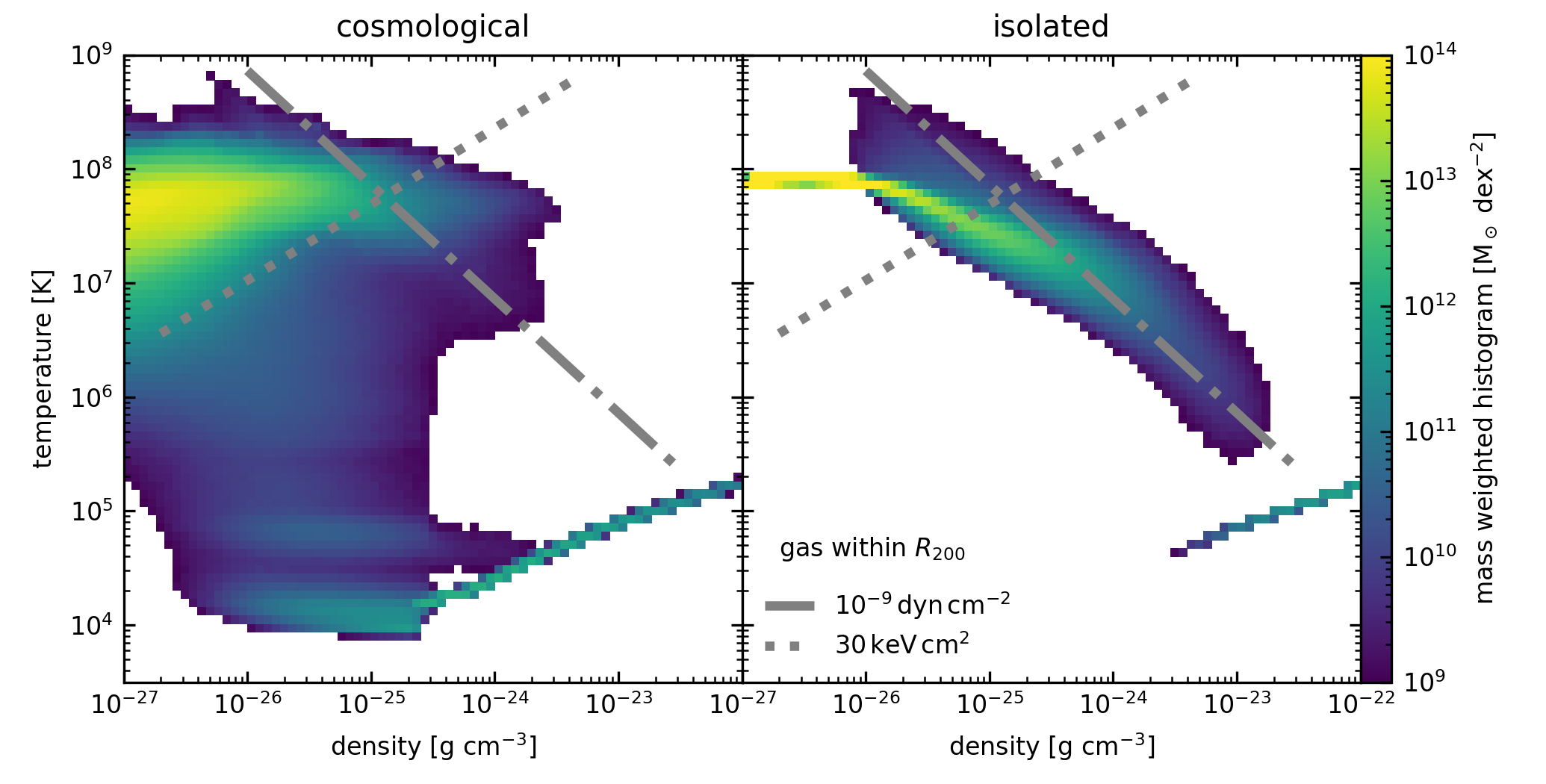}
    \includegraphics[]{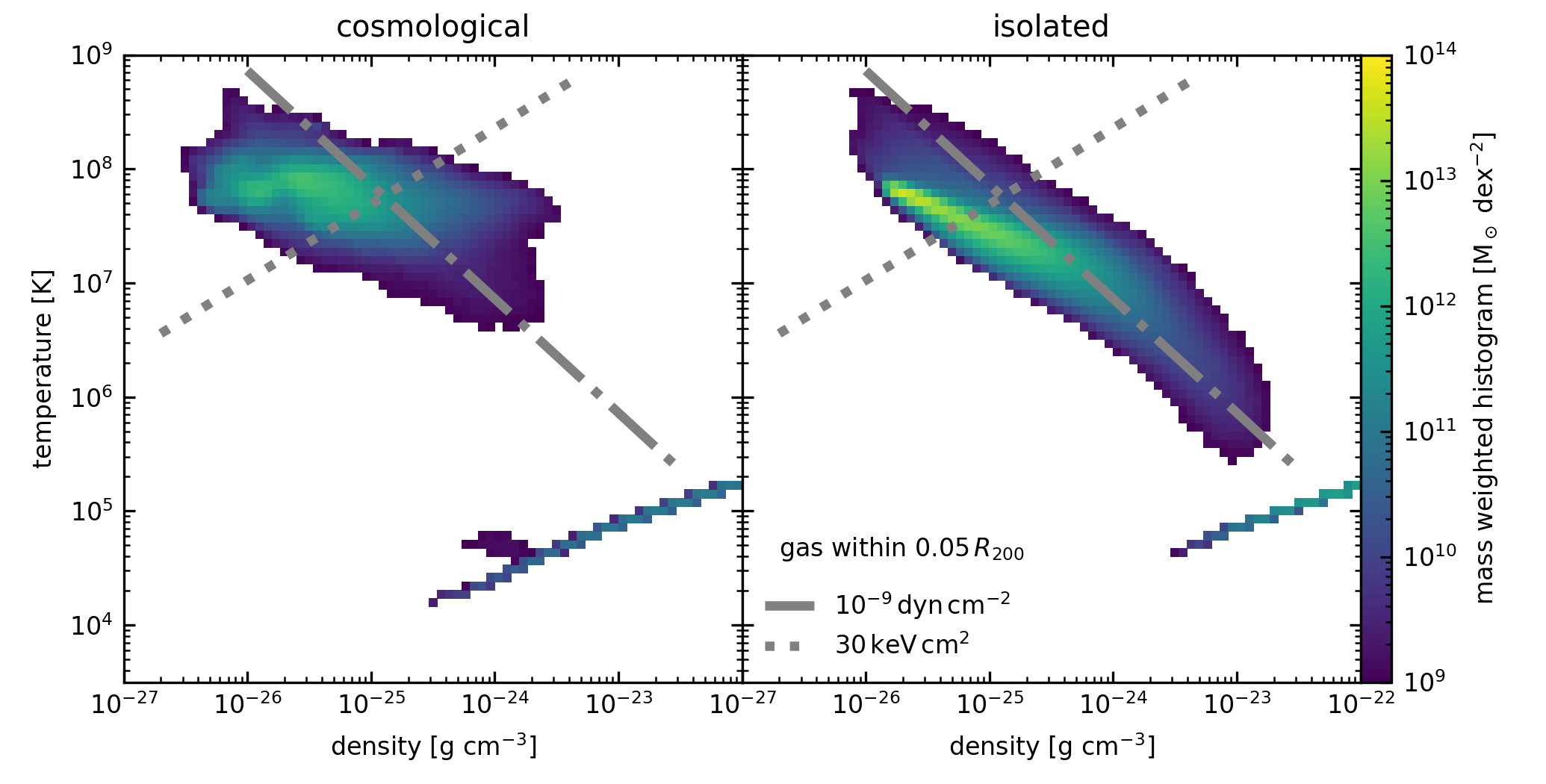}
    \caption{Top: phase diagram of the simulated ICM for gas within $R_{200}$.
    Bottom: corresponding phase diagrams restricted to gas within $0.05\,R_{200}$.
    Left panels show the stacked cosmological simulations (3 halos), while the right panels display the isolated simulation. All are a time-average over $10\,\mathrm{Gyr}$. The grey dotted and dash-dotted lines indicate an adiabat and isobar, respectively, to facilitate comparison between panels. Warm gas with $T < 10^7 \,{\rm K}$ is found outside the central region in cosmological simulations but not in the isolated simulation.}
\label{fig:phase_diagram_cosmo_isolated}
\end{figure*}

Having considered the radially averaged distribution of the ICM, as well as the global time evolution of the galaxy cluster, we now focus on the temperature--density phase diagram of the gas in Fig.~\ref{fig:phase_diagram_cosmo_isolated}. The left panels show the stack of all snapshots between $z=2$ and $z=0$ of the three cosmological halos, the right panels the stack of the corresponding snapshots in the isolated simulation. The top row shows all gas within $R_{200}$, while the bottom shows only gas in the core, within $0.05\,R_{200}$. Note that we cap the two-dimensional histogram at the lower limit of the colour map, which hides potential features due to jets that make up a comparably small amount of mass. The hot, low density ICM shows a sharp feature in the isolated simulations, while in the cosmological simulation, the temperature distribution is generally broader, likely due to the halo growth and consequently an increase in virial temperature over time as well as due to substructure. The isolated cluster reaches higher densities and lower temperatures in the ICM, consistent with the radial profiles. The diagonal feature in the dense, low temperature part is the effective equation of state model for the unresolved interstellar medium \citep{Springel2003} and by construction at the same location, though not necessarily populated in the same way, in all simulations. Most prominently, the cosmological simulations show a considerable amount of sub-viral ($<10^{7}\,\mathrm{K}$) gas at low density. Comparing the histograms for all gas and only for gas constrained to radii smaller than $0.05\,R_{200}$, it is evident that this gas is located at considerable distance from the centre, with all of the colder gas in the centre being star forming (i.e. having a density exceeding the star formation threshold). This is consistent with the lower two panels in Fig.~\ref{fig:mass_time}, where the cosmological simulations show between $10^{10}\,\mathrm{M}_\odot$ and $10^{11}\,\mathrm{M}_\odot$ of warm, non-star-forming gas while barely any of this is present in the isolated simulation. In contrast, the star forming gas mass is comparable between the two. Overall, this points to a cosmological origin for the warm gas beyond the core radius. 

\section{Discussion}
\label{sec:discussion}

In this work, we use three cosmological and one isolated galaxy cluster simulations \citep{Ehlert2023} to identify the impact of jet feedback during cosmological assembly on ICM properties. To ensure that this is a meaningful comparison, we verify that the global baryonic content is in agreement with observations (Fig.~\ref{fig:masses_multiplot}), and that the ICM thermodynamic properties between cosmological simulations at $z=0$ and the isolated halo after a $1\,\mathrm{Gyr}$ initialization period are comparable. We do the latter via projections (Fig.~\ref{fig:projections}) and radial profiles (Fig.~\ref{fig:profiles}), and find good agreement with each other and with a fit to the Perseus cluster electron density and temperature profiles \citep{Churazov2003}.

\subsection{Cluster turbulence}

The velocity dispersion projections of the cluster core show substantially lower degrees of intra-cluster turbulence throughout the isolated cluster's atmosphere (Fig.~\ref{fig:projections_core}). This is unsurprising at larger radii beyond the reach of the AGN jet. However, we find that the core region, defined here as $0.05\, R_{200}\approx 100\,\mathrm{kpc}$, to be affected as well. Using the filtering approach of \citet{Perrone2026}, we find that the central AGN jet feedback drives the non-thermal pressure fraction to the few to ten percent level in isolated and cosmological setups only in the inner few tens of kpc. Turbulent pressure support at larger radii lies at a few percent (Fig.~\ref{fig:nt_profiles}), in agreement with recent observational studies \citep{Zhang2025, TheXrismCollaboration2026}, and is driven to a large degree by cosmological infall and interaction processes of satellite galaxies with the ICM. This behaviour is consistent with a recent study comparing individual galaxy cluster mergers and periodic (but not self-regulated) jet episodes \citep{Bellomi2026}, and is qualitatively consistent with previous work \citep{Bourne2017}. In this work, we show this mechanism also operates in more complex setups with a self-regulated heating--cooling cycle and a self-consistent cosmological formation history over cosmic time. Interestingly, the degree of turbulence is somewhat larger than what is commonly found in simulations of single jet outbursts \citep{Ehlert2021, Li2025}. We speculate that this discrepancy is due to a combination of radiative cooling and self-regulated jet feedback. More detailed studies of jet-ICM interaction in the presence or absence of cooling flows are needed to resolve this question.

\subsection{Thermodynamic profiles and AGN feedback self-regulation}

The thermodynamic profiles in both cosmologically-forming and isolated halos exhibit a cool-core structure (Fig.~\ref{fig:profiles}). This is in stark contrast with PICO-Cluster simulations of the same halos using the IllustrisTNG model, which predicts that all three halos are non-cool-core clusters. This discrepancy highlights the influence of SMBH accretion and AGN feedback coupling with the surrounding ICM on the thermodynamic state in the cluster core. We attribute the differences in central entropy to two effects: a dependence of central entropy on SMBH mass, and differences in how feedback energy couples to the central gas. Regarding the former, Fig.~\ref{fig:masses_multiplot} shows that the IllustrisTNG model produces SMBH masses a factor of $2$--$3$ larger, which, via Eq.~(\ref{eq:entropy}), translates to central entropies a factor of $2.5$--$4$ higher. This alone would raise central entropies from ${\sim} 15\,{\rm keV\,cm}^2$ to ${\sim} 50\,{\rm keV\,cm}^2$, pushing them beyond the cool-core threshold, yet still well below the $100$--$250\,{\rm keV\,cm}^2$ levels seen in Fig.~\ref{fig:profiles}. The remaining discrepancy is likely due to the latter effect.

For simulations with the jet model, the isolated halo has a more pronounced cool-core. This is largely due to the higher SMBH masses. Figure~\ref{fig:mass_time} shows that while the SMBH mass growth in the isolated simulations is small (for the adopted efficiencies), it increases by about an order of magnitude in the cosmological simulations. \citet{Weinberger2018} showed that at these high SMBH masses, the main growth channel is hierarchical merging with other SMBHs, which is absent in isolated simulations, leading to an indirect impact on central entropy profiles. Note that this effect might be exaggerated in our model due to the assumption of instantaneous merging of close SMBHs \citep{Weinberger2017}. Future simulations using a more detailed modelling of SMBH dynamics \citep[similar to][]{Tremmel2019, Ni2022, Bhowmick2025} will likely lead to more off-centre SMBHs in the halo, fewer mergers, and lower central SMBH masses in clusters. It remains to be investigated whether such a lowering of SMBH masses in cosmological setups leads to similar levels of central entropy as in the isolated setup, or if other effects such as mergers prevent this from happening.

The energetics for both cosmological and isolated halos are predominantly dictated by a balance of cooling and AGN jet heating on Gyr timescales. This reduces the central star formation to comparably small values (Fig.~\ref{fig:sfr_energy_rate_time}). The isolated halo shows slightly higher levels of star formation in the central galaxy, possibly related to the lower-entropy core. The cosmological halos in contrast, while having considerable in-situ star formation in the central galaxy at $z=0.5-2$, see a stronger reduction of star formation in the centre. However, when measured in the entire halo, there is only a mild redshift dependence of star formation.  Overall we conclude that both isolated and cosmological halos reach a similar state of self-regulation, i.e., self-regulation is an equilibrium solution intrinsic to this system. The IllustrisTNG simulations show the most significant deviations form the jet model in the in-situ star formation of the central galaxy. We emphasise at this point that our simulations model the gas as an ideal radiative fluid with heating via hydrodynamical AGN jets. Non-thermal components such as magnetic fields, cosmic rays, as well as non-ideal fluid effects such as conduction or viscosity are not modelled in the jet simulations. We find these effects are not necessary to establish the basic cool-core state from cosmological initial conditions, but it is likely that they modify said state. The setup presented in this work can provide the basis for future studies that investigate the question of missing physics in detail.

\subsection{Warm gas in the ICM}

A significant difference between the cosmological and the isolated halos is the thermodynamic state of the gas in the halo (Fig.~\ref{fig:phase_diagram_cosmo_isolated}, top panels). While the isolated halo contains dilute gas ($\rho<10^{-25}$g~cm$^{-3}$) only in the hot phase, the cosmological halos show substantial amounts of comparatively low-density, lower temperature gas. This gas is primarily located at significant distances from the halo centre (i.e., absent in the bottom left panel of Fig.~\ref{fig:phase_diagram_cosmo_isolated}), and clearly associated with cosmological assembly. This could be either due to gas brought in directly through accretion, or due to formation induced by the cosmological assembly process. In contrast, the isolated simulations shows a maximum radius for cold gas which can be interpreted as the reach of AGN feedback. This finding is qualitatively similar to the study of \citet{Fielding2020} who compared a number of different isolated and cosmological simulations of Milky-Way mass halos and found a different origin of cold gas in the inner and the outer halo with a dividing line at around $0.5~R_{200}$. For galaxy clusters, we find this dividing line to be around $100$~kpc or $0.05~R_{200}$. This can be qualitatively understood by considering that the cumulative AGN feedback energy, which, neglecting mergers and assuming a constant feedback efficiency, is proportional to the black hole mass, 
\begin{align}
  E_{\rm AGN}\sim \epsilon M_{\bullet} c^2. 
\end{align}
The gravitational binding energy of the ICM gas scales as 
\begin{align}
    E_{\rm gas} \sim \frac{G M_{200} M_{\rm gas}}{R_{200}} \propto f_{\rm gas} M_{200}^{5/3} \rho_c^{1/3}.
\end{align}
Assuming a correlation $M_\bullet \propto M_{200}^\alpha$ with $\alpha < 5/3$, it is clear that the ratio $E_{\rm AGN}/E_{\rm gas}$ decreases with halo mass, implying a  shorter reach in more massive halos. Alternative arguments based on entropy of the feedback-heated gas come to the same conclusion \citep{Pontzen2026}. 

While the infalling gas in colder phases is a subdominant part of the overall ICM mass budget, it still adds up to about $(10^{10}$--$10^{11})\,\mathrm{M}_\odot$ (Fig.~\ref{fig:mass_time}) and thus may be important for fuelling star formation in cluster galaxies at levels of $(10$--$100)\,\mathrm{M}_\odot\,\mathrm{yr}^{-1}$ (Fig.~\ref{fig:sfr_energy_rate_time}).

\subsection{Limitations and future work}
We note however that due to insufficient resolution, the dynamics of this cooler gas is not faithfully followed in the simulations presented here. The cooling length of $10^5\,\mathrm{K}$ gas even at comparably low densities drops significantly below kpc scales, making it computationally costly to resolve the collapse even to the warm phase \citep[see e.g.][for significantly lower mass halos]{vandeVoort2019}, a challenge that requires dedicated simulation techniques \citep[e.g.][]{Hummels2019, Ramesh2024}. The long-term evolution, governed by the efficiency of surface effects like radiative mixing layers and shattering, poses even more restrictive resolution requirements \citep{Ramesh2026} making such approaches infeasible over global timescales.
Thus, questions about the fate of the warm gas, whether it might rain onto the central galaxy, form stars in the halo, or get mixed into the hot ICM, will be subject to future studies that address the above mentioned computational challenges. This can be done by incorporating recently developed multi-fluid techniques \citep{Weinberger2023, Das2024, Bollati2026} in this setup.

Another important limitation of this work is the fidelity of the jet modelling in the simulations. The two main areas of concern are black hole trajectory and jet resolution. The former can be addressed using live dynamical friction treatments \citep[e.g.][]{Tremmel2015, Pfister2019, Ma2023, Genina2024}, the latter requires an increase in jet resolution and is well-studied for this model \citep{Weinberger2023b}. While the spatial resolution requirements seem feasible with adaptive refinement, especially compared to the typical resolution in the ISM gas, the key problem is that the jet material has the highest velocity and sound speed in the simulation. This naturally leads to a Courant-Friedrichs-Lewy timestep criterion for jet cells that is far shorter than that of the surrounding medium. In practice, this means that the simulations presented require tens of millions of timesteps to reach $z=0$ which is reaching the limits of what is currently possible. Increasing the spatial resolution would increase the number of timesteps by at least the same amount (likely slightly exceeding it since the tail of the velocity distribution of cells is better sampled). A possible solution to this problem subject to future work is to run only parts of the simulation, e.g. toward $z=0$ at this high resolution. Similar efforts have been done before \citep{Bourne2019,Yates-Jones2023}, but usually switching the AGN feedback models between low and high-resolution phases. With the set of simulations presented in this work, we do not require such a switch, merely a change in jet resolution (i.e., a single parameter), which will likely lead to a more accurate modelling of the interaction between large and small scales effects. 

\section{Conclusion}
\label{sec:conclusion}

We have presented four hydrodynamical simulations of cool-core galaxy clusters performed with the moving-mesh code \textsc{Arepo}, using the AGN driven jet feedback from \citet{Weinberger2023b} and the IllustrisTNG galaxy formation model \citep{Weinberger2017, Pillepich2018}. Three cosmological zoom simulations of massive halos were specifically selected from the PICO-Cluster initial conditions \citep{Berlok2026} to achieve the lowest possible central cluster entropy state. The fourth was the isolated galaxy cluster presented in \citet{Ehlert2023}, set up as a hydrostatic gas halo balancing an analytic gravitational potential. By comparing the simulated intra-cluster gas of the different halos, we study the role of cosmological hierarchical assembly on the intra-cluster medium properties of cool-core galaxy clusters. By comparing with fiducial IllustrisTNG model simulations, we assess the importance of the specific AGN feedback model. The key findings are:

\begin{enumerate}
    \item Density, temperature and cluster entropy profiles of our AGN jet feedback model resemble the Perseus cluster. This state is achievable both with isolated and cosmological initial conditions. We find the model predicting the central entropy by a heating-cooling equilibrium that non-linearly depends on the SMBH mass \citep{Weinberger2026} to be an excellent match for our simulations. This implies that cosmological SMBH growth via mergers has a significant impact on central entropies.
    \item For the same initial conditions, the PICO-Cluster reference simulations performed with the IllustrisTNG model produce non-cool-core clusters. This implies that the central entropy levels in the simulated halos are mostly dictated by model choices, most notably in the SMBH growth history and the AGN feedback model, rather than by cosmological environment. While the kinetic feedback mode of IllustrisTNG creates shocks that quickly dissipate the feedback energy \citep[][their Fig. 1-3]{Weinberger2017}, the lobes inflated by low-density AGN jets rise buoyantly and merely drive subsonic turbulence while mixing different temperature fluids \citep[][their Fig. 10]{Meenakshi2026}, thereby preserving the cool-core state without increasing its entropy beyond the threshold for non-cool-core clusters.
    \item Self-regulation is achieved in both isolated and cosmological halos, with star formation being the residual of an approximate time-averaged heating-cooling balance. While the central star formation rates in the isolated setup tend to be larger than in the cosmological halos, the overall star formation rate in the halo (including satellites) is comparable.
    \item Comparing cosmological to isolated simulations using spatial filtering at fixed length \citep{Perrone2026}, we find that the velocity structure of the ICM gas at radii $>50\,\mathrm{kpc}$ is mostly driven by cosmological formation, most likely mergers with other halos \citep{Bellomi2026}, with possible contributions from internal processes in satellite galaxies. Central, self-regulated AGN jet heating only contributes to the observed levels of turbulence in the centre.
    \item Warm (${<}10^5\,{\rm K}$) gas beyond the core ($0.05\,R_{200\mathrm{c}}$) is only present in cosmological halos. While we cannot reliably predict the long-term behaviour of this cool gas, we speculate that it has substantial implications for related observables such as H$\alpha$ emission. 
\end{enumerate}

These findings show that while it might not be necessary to use cosmological simulations to establish the feedback cycle in galaxy clusters, the cosmological assembly is crucial to understand the observable velocity structure of X-ray emitting gas and the abundance and kinematics of warm gas in cluster atmospheres. Using cosmological simulations we obtain different central gas profiles from runs starting from the same initial conditions but using different AGN feedback models. This highlights the importance of an accurate AGN feedback model to study galaxy cluster cores. The employment of the computationally more expensive jet model might, however, be less crucial for aspects that are less sensitive to the effects of AGN feedback. 

With this work, we have laid the groundwork to better model AGN jet-ICM interactions in a cosmological context; yet more, higher resolution simulations are needed to better understand the details of jet propagation in these realistic environments. On top of this, further research is required to study the long-term behaviour of warm and cold gas  phases, and to extend the sample to lower mass halos.

\begin{acknowledgements}

RW acknowledges funding of a Leibniz Junior Research Group (project number J131/2022). CP and LJ acknowledge support by the Deutsche Forschungsgemeinschaft (German Research Foundation) for the Research Unit FOR5195 on Relativistic Jets in Active Galaxies (443220636). CP, LMP and JW acknowledge support by the European Research Council under ERC-AdG grant PICOGAL-101019746. TB acknowledges financial support by the Carlsberg Foundation via grant CF23-0417. The authors gratefully acknowledge the computing time granted by the Resource Allocation Board and provided on SuperMUC-NG through the project ``The plasma physics of galaxy clusters in a cosmological context'' (ID: pn68cu).
\end{acknowledgements}

\bibliographystyle{aa}

\begin{appendix}    

\section{Resolution and time dependence in the isolated setup}
\label{sec:appendix_res}

\begin{figure}
    \centering
    \includegraphics[width=1.0\linewidth]{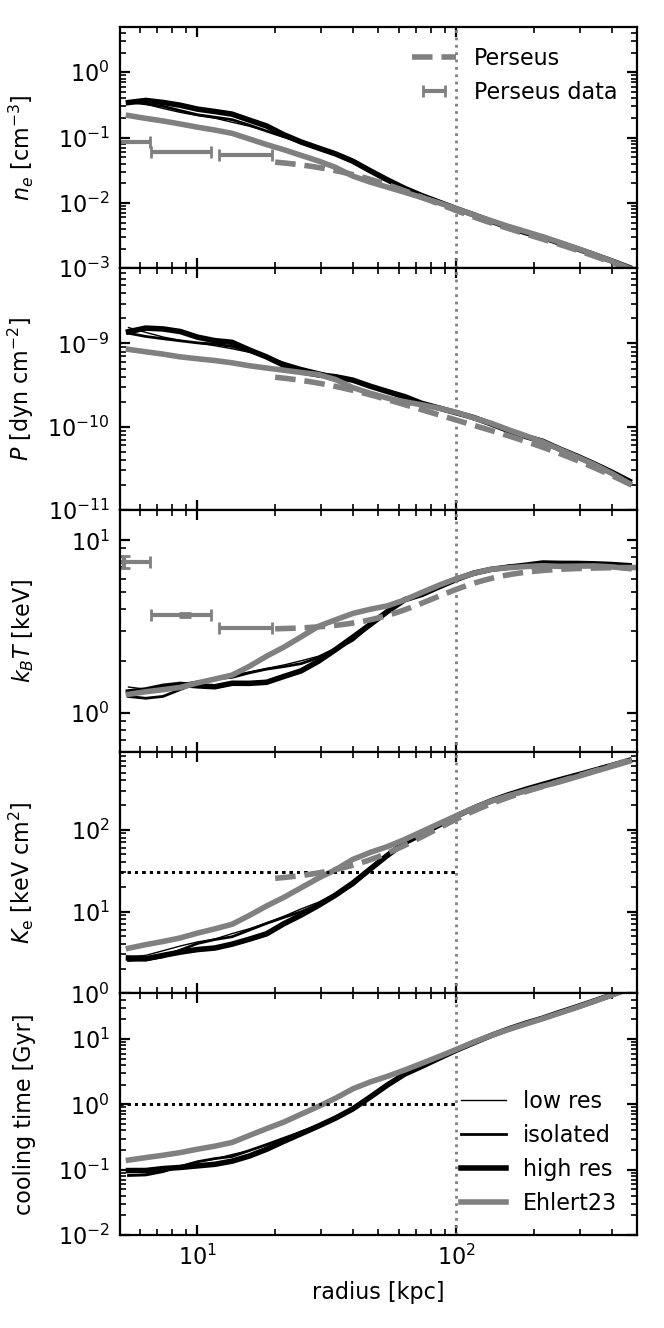}
    \caption{Thermodynamic profiles as in Fig.~\ref{fig:profiles}, showing the isolated simulation with different jet resolutions, with the intermediate resolution being the one shown in the main part of the paper, while low- and high-res simulations denote runs with different $V_{\rm jet, target}$ but otherwise identical settings. For reference, we include the thermodynamic profiles of the high-resolution magneto-hydrodynamics, Bondi-accretion run from \citet{Ehlert2023}.}
    \label{fig:profiles_res}
\end{figure}

\begin{figure}
    \centering
    \includegraphics[width=1.0\linewidth]{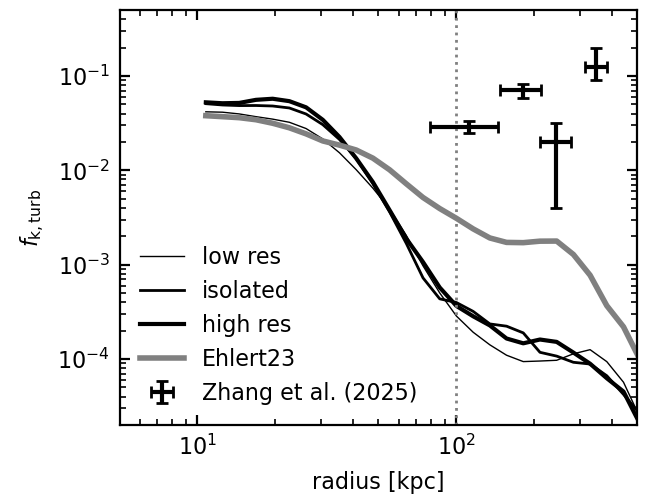}
    \caption{Nonthermal pressure fraction similar in Fig.~\ref{fig:nt_profiles} with spatial filtering at $\sigma_{\ell} = 20$~kpc,
    showing the isolated simulation with different jet resolutions. For reference, we include the the high-resolution magneto-hydrodynamics, Bondi-accretion run from \citet{Ehlert2023}.}
    \label{fig:nt_profiles_res}
\end{figure}

\begin{figure}
    \centering
    \includegraphics[width=1.0\linewidth]{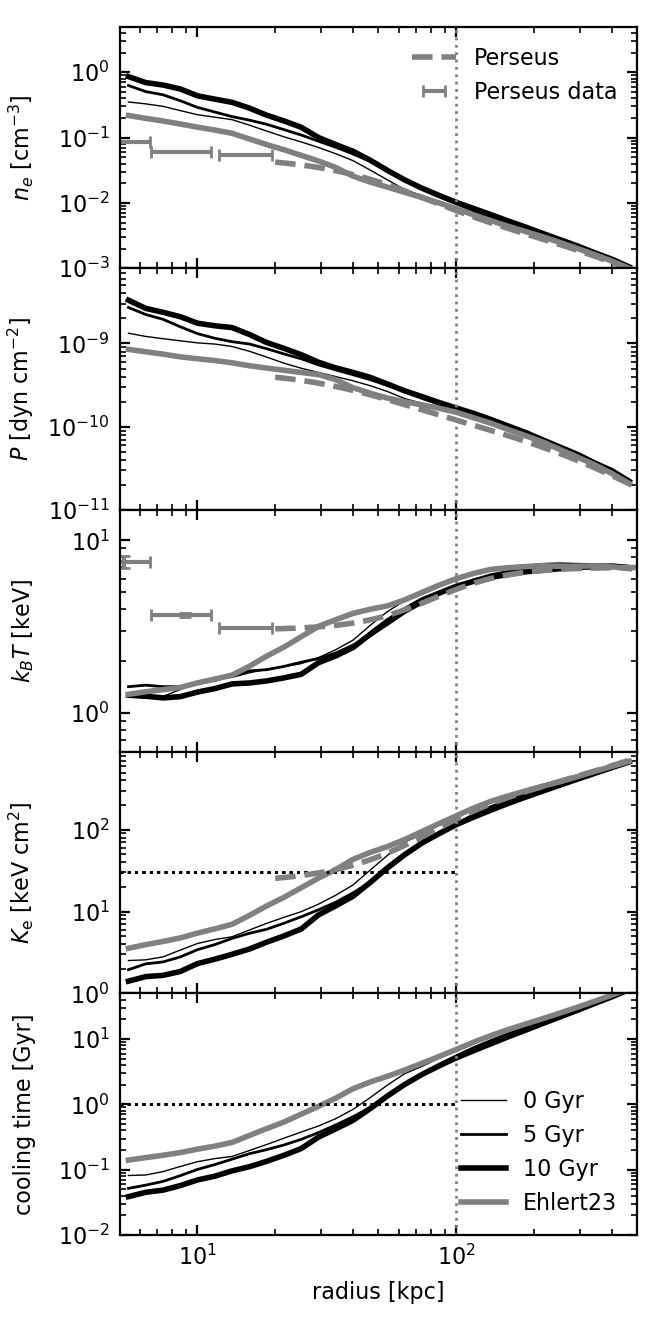}
    \caption{Thermodynamic profiles as in Fig.~\ref{fig:profiles}, showing the isolated simulation at three different times. For reference, we also include the thermodynamic profiles of the high-resolution magneto-hydrodynamics, Bondi-accretion run from \citet{Ehlert2023} at $1\,{\rm Gyr}$.}
    \label{fig:profiles_time}
\end{figure}

\begin{figure}
    \centering
    \includegraphics[width=1.0\linewidth]{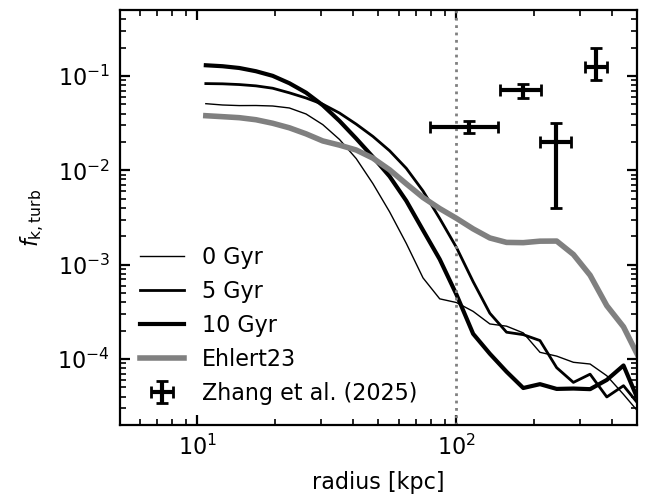}
    \caption{Nonthermal pressure fraction similar in Fig.~\ref{fig:nt_profiles} with spatial filtering at  $\sigma_{\ell} = 20$~kpc,
    showing the isolated simulation at three different times. For reference, we also include the high-resolution magneto-hydrodynamics, Bondi-accretion run from \citet{Ehlert2023}.}
    \label{fig:nt_profiles_time}
\end{figure}

One of the key aspects of AGN jet feedback modelling is the inability to fully capture the dynamic range of this phenomenon. Thus, it is key to carefully consider if the employed resolution is sufficient for the intended study. In \citet{Weinberger2023b}, we showed that different aspects of jets require different resolutions: jet propagation has significantly higher resolution requirements than accurately capturing the effect of a single jet outburst on the ICM. Indeed, the employed resolution in our cosmological simulations of $V_\textrm{jet, target}^{1/3} = 1.39 \,\mathrm{kpc}$ is slightly coarser than even the lowest resolution simulation \texttt{rw jet a 1} of \citet{Weinberger2023b}, and a factor $10$ larger target volume than the fiducial runs of \citet{Ehlert2023}. The isolated simulation has a jet resolution of $V_\textrm{jet, target}^{1/3} = 0.65 \,\mathrm{kpc}$, like the fiducial simulations of \citet{Ehlert2023} and is comparable to the resolution of simulation \texttt{rw jet a 2} in \citet{Weinberger2023b}. Consequently, the jet propagation is certainly not adequately resolved in either of the simulations.
Regarding the effect of a given jet outburst on the ICM, the resolution in the isolated system leads to converged results, while the coarser resolution used in cosmological simulations leads to moderate resolution effects \citep[][their Fig.~5]{Weinberger2023b}. How these jet resolutions affect thermodynamic profiles in self-regulated setups is shown in Fig.~\ref{fig:profiles_res}. We compare simulations with the isolated setup at low resolution (equivalent to the cosmological simulations), intermediate resolution (that of the isolated simulation discussed throughout the paper) and high resolution (corresponding to the high-resolution simulations of \citealt{Ehlert2023}). For reference, we also show the high-resolution Bondi simulation of \citet{Ehlert2023}. Note that the latter simulation contains magnetic fields, while our resolution study is purely hydrodynamic, which allows us to compare the resolution effects to physics changes. Overall, we conclude that even the lowest resolution jets are adequate for studying the thermodynamic state of the ICM, and differences discussed in this paper are not a consequence of inadequate resolution. Figure~\ref{fig:nt_profiles_res} shows the respective non-thermal pressure using Eq.~(\ref{eq:fkturb}), which is computed using spatial filtering at fixed length \citep[effective spread $\sigma_\ell = 20\,\mathrm{kpc}$, see][]{Perrone2026}. While the isolated simulation are converged, the presence of an initial magnetic field increases the level of turbulence in the \citet{Ehlert2023} run, however not sufficiently to explain observed levels.

Figure~\ref{fig:profiles_time} shows the thermodynamic profiles at different times, $0$, $5$ and $10\,\mathrm{Gyr}$. Note that $0\,\mathrm{Gyr}$ is still after an initialization period of $1\,\mathrm{Gyr}$, i.e., it does not represent the initial conditions but an equilibrium state. Even after $5\,\mathrm{Gyr}$, the thermodynamic profiles are practically unchanged. After $10\,\mathrm{Gyr}$, the central ICM density has increased somewhat, and the temperature has decreased. As this work shows, however, an isolated setup is increasingly unrealistic at these timescales, as mergers and ram-pressure stripping become increasingly important. Thus, we conclude that the profiles of our isolated simulations are very stable also over timescales of several Gyrs. Figure~\ref{fig:nt_profiles_time} shows the respective non-thermal pressure profiles using spatial filtering at effective spread $\sigma_\ell = 20\,\mathrm{kpc}$. While there is some time dependence, even after $10~{\rm Gyr}$ of evolution, the level of turbulence in the outskirts remains low.

\label{LastPage}

\end{appendix}

\end{document}